\documentclass[a4paper,twocolumn,11pt,accepted=2026-06-03]{quantumarticle}
\pdfoutput=1
\usepackage[utf8]{inputenc}
\usepackage[english]{babel}
\usepackage[T1]{fontenc}
\usepackage{amsmath}
\usepackage{hyperref}
\usepackage[numbers]{natbib}
\usepackage{amssymb}
\usepackage{braket}
\usepackage{commath}
\usepackage{mathrsfs}
\usepackage{caption}
\usepackage{subcaption}
\usepackage{tikz}
\usepackage{lipsum}
\usepackage{booktabs}

\def\ddefloop#1{\ifx\ddefloop#1\else\ddef{#1}\expandafter\ddefloop\fi}

\def\ddef#1{\expandafter\def\csname bf#1\endcsname{\ensuremath{\mathbf{#1}}}}
\ddefloop ABCDEFGHIJKLMNOPQRSTUVWXYZabcdefghijklmnopqrstuvwxyz0\ddefloop

\def\ddef#1{\expandafter\def\csname v#1\endcsname{\ensuremath{\boldsymbol{#1}}}}
\ddefloop ABCDEFGHIJKLMNOPQRSTUVWXYZabcdefghijklmnopqrstuvwxyz\ddefloop

\def\ddef#1{\expandafter\def\csname v#1\endcsname{\ensuremath{\boldsymbol{\csname #1\endcsname}}}}
\ddefloop {alpha}{beta}{gamma}{delta}{epsilon}{varepsilon}{zeta}{eta}{theta}{vartheta}{iota}{kappa}{lambda}{mu}{nu}{xi}{pi}{varpi}{rho}{varrho}{sigma}{varsigma}{tau}{upsilon}{phi}{varphi}{chi}{psi}{omega}{Gamma}{Delta}{Theta}{Lambda}{Xi}{Pi}{Sigma}{varSigma}{Upsilon}{Phi}{Psi}{Omega}{ell}\ddefloop

\DeclareMathOperator{\im}{\mathrm{i}}

\begin{document}

\title{Learning thermodynamic master equations for open quantum systems}%

\author{Peter Sentz}
\affiliation{Department of Data and Decision Sciences, Emory University, Atlanta, GA  30322}
\email{peter.thomas.sentz@emory.edu}
\author{Stanley Nicholson}
\affiliation{Division of Applied Mathematics, Brown University, Providence, RI  02912}
\author{Yujin Cho}
\affiliation{Lawrence Livermore National Laboratory, Livermore, CA  94550}
\author{Sohail Reddy}
\affiliation{Lawrence Livermore National Laboratory, Livermore, CA  94550}
\author{Brendan Keith}
\affiliation{Division of Applied Mathematics, Brown University, Providence, RI  02912}

\author{Stefanie G{\"u}nther}%
\affiliation{Lawrence Livermore National Laboratory, Livermore, CA  94550}

\maketitle

\begin{abstract}
The characterization of Hamiltonians and other components of open quantum dynamical systems plays a crucial role in quantum computing and other applications. Scientific machine learning techniques have been applied to this problem in a variety of ways, including by modeling with deep neural networks. However, the majority of mathematical models describing open quantum systems are linear, and the natural nonlinearities in learnable models have not been incorporated using physical principles. We present a data-driven model for open quantum systems that includes learnable, thermodynamically consistent terms. The trained model is interpretable, as it directly estimates the system Hamiltonian and linear components of coupling to the environment. We validate the model on synthetic two and three-level data, as well as experimental two-level data collected from a quantum device at Lawrence Livermore National Laboratory.
\end{abstract}

\section{\label{sec:intro}Introduction}
Characterizing quantum dynamical processes is central to the advancement of quantum computing devices, with significant impacts on quantum control, compilation, and error mitigation \cite{corcoles2019challenges}.  
Accurate dynamical models are essential to address two challenges: 1) optimized quantum compilation, enabling optimal quantum gate sequences and resource allocation, and 2) quantum error mitigation and correction to achieve overall improved system performance \cite{tuckett2019tailoring}.
A precise model of the underlying quantum dynamics requires accurate knowledge of system parameters such as qubit frequencies, coupling strengths, and noise channels.
Extracting such information from measurement data is inherently challenging, yet recent advances in data-driven methods incorporating machine learning (ML) have led to significant progress in this area. A comprehensive overview of ML for quantum systems can be found in \cite{gebhart2023learning}, featuring recent developments for the tasks of learning quantum states, quantum dynamics, and quantum measurements.

Hamiltonian dynamics can be reconstructed using observed time series of a complete set of measurements~\cite{baldwin2014quantum}, and quantum process tomography allows for the identification of linear non-unitary dynamics governed by the Lindblad master equations~\cite{nielsen2010quantum}. However, the number of measurements required to fully characterize the dynamics of a quantum system grows exponentially with the number of qubits. Therefore, a variety of approaches have been developed that pursue estimates of the unknown parameters of the Hamiltonian and Lindblad operators using limited measurements~\cite{cerrillo2014non, samach2022lindblad, huang2023learning, hangleiter2024robustly}. Nevertheless, a growing array of approaches incorporate deep neural networks in the characterization process, introducing nonlinear transformations of the quantum state into the dynamical model~\cite{mohseni2024deep, zhu2023quantum, lewis2024improved, liu2022solving, hartmann2019neural}. 

For instance, recurrent neural networks (RNNs) have been employed to predict measurement outcomes under dissipative dynamics \cite{leclerc2021predicting}, and to reconstruct quantum dynamics using trajectory data from superconducting qubits under microwave control sequences \cite{flurin2020using}. These methods, which leverage machine learning to approximate quantum states and dynamical processes, often fall under the category of ``black-box'' approaches, where the learned model lacks explicit physical interpretability, hence requiring post-processing of the learned agent.

On the other hand, ``gray-box'' approaches combine data-driven learning models with physically understandable dynamical structures, incorporating known physics directly into the machine learning agent to improve accuracy and interpretability. 
For example, Youssry et al.\ \cite{youssry2020characterization} combine the output of a black-box machine learning agent with numerically evolved quantum dynamical processes to address uncertainties in the Hamiltonian model and capture noise characteristics. 

More intrusively, machine learning models based on physics-informed neural ordinary differential equations \cite{lai2021structural}, also termed universal differential equations (UDEs) \cite{rackauckas2020universal}, use trajectory data to train a correction to an underlying approximate physical model described by a differential equation. Such learning approaches often require smaller training data sets to achieve accurate model fidelity and are more resilient to noisy training data~\cite{banchi2018modelling, krastanov2020unboxing}. The UDE approach was recently applied to the problem of Hamiltonian learning in~\cite{heightman2024solving}, where a nonlinear correction to the Hamiltonian dynamics is learned from trajectory data.

An application of UDEs to learning open quantum systems is proposed in~\cite{reddy2024data}, where a parameterized augmentation of the Lindblad equation was trained to identify latent dynamics in a single qubit superconducting quantum processing unit using microwave-driven control pulses. There, a structure-preserving ansatz to learn corrections to approximate Hamiltonian and Lindblad decoherence operators is compared to a nonlinear augmentation using a nonlinear feed-forward neural network. While the former guarantees completely-positive trace-preserving (CPTP) maps~\cite{nielsen2010quantum}, the latter has demonstrated higher model accuracy after training, indicating the need for nonlinear augmentations to the linear Lindblad equation. However, ensuring that a learned nonlinear dynamical model yields physically consistent dynamics when evaluated outside of the training data remains a crucial challenge. 

Approaches that incorporate physical laws as inductive biases~\cite{cuomo2022scientific, faroughi2022physics, huang2025partial, kim2021knowledge}, have demonstrated improvements in both accuracy and generalization. Thus, incorporating nonlinearity into quantum master equations in a physically consistent manner has the potential for more robust learned models. In particular, a recent family of ML techniques that embed the laws of thermodynamics into the learnable model has been successfully applied to dissipative dynamical systems~\cite{cueto2023thermodynamics}. Among these methods, the so-called metriplectic~\cite{morrison1986paradigm} neural networks are based on a unifying formalism for reversible and irreversible dynamics known as the generic equation for non-equilibrium reversible-irreversible coupling (GENERIC)~\cite{grmela1997dynamics, ottinger1997dynamics, ottinger2005beyond}. Thermodynamics-informed neural networks (TINNs) parametrize the GENERIC structure of dynamics as neural networks, and weakly enforce the necessary symmetry and degeneracy conditions through penalization terms in the loss function~\cite{hernandez2021deep, chinesta2020learning, hernandez2022thermodynamics}. In contrast, GENERIC formalism-informed neural networks (GFINNs) employ architectures that directly enforce the properties of GENERIC for every configuration of the neural network parameters~\cite{zhang2022gfinns}. Other works strongly enforce the necessary conditions by learning structure-preserving brackets parametrized by learnable tensors~\cite{lee2021machine, gruber2023reversible}. The recent port-metriplectic approach~\cite{hernandez2023port} decomposes dissipative systems into local subsystems, reducing the computational complexity of training the model. In this work, we extend the use of ML enforcing thermodynamic consistency to the context of open quantum dynamical systems.

\subsection{Contributions and highlights}
We present a thermodynamically consistent, data-driven model for learning dynamics of open quantum systems. In particular, the model incorporates nonlinear terms based on physical principles via the GENERIC formalism, in contrast to methods that augment linear master equations with nonlinear transformations in an ad hoc manner. This yields dynamical equations that guarantee valid quantum states for two-level systems. The learned model is interpretable, directly providing an estimate of the Hamiltonian and the operators that govern the coupling to an external environment. We describe this model within the density operator formalism and detail its construction using the density operator's representation in the Pauli or Gell--Mann basis.

Next, we test the accuracy and generalizability of the model on two (qubit) and three (qutrit) level synthetic quantum systems. Finally, we validate its efficiency on noisy, two-level experimental data taken from Lawrence Livermore National Laboratory's (LLNL) QuDIT device.

\subsection{Outline}

We continue by discussing the shortcomings of the linear Lindblad master equation in terms of thermodynamical consistency and introduce an alternative nonlinear master equation (Section \ref{sec:dissipative_sys}). We then reformulate the master equation in terms of generalized Bloch vector dynamics in Section \ref{sec:blochvector_dynamics}, and show in Section \ref{sec:nn_dynamics} how this model can be augmented or approximated by neural networks, addressing key features such as Hamiltonian characterization, environmental coupling operators, and nonlinear interactions. Our numerical results in Section \ref{sec:numerics} demonstrate the effectiveness of the proposed approach on synthetic and experimental data. Conclusions and future work are discussed in Section \ref{sec:conclusion}.

\section{\label{sec:dissipative_sys}Nonlinear thermodynamic master equation}
The dynamics of open quantum systems are conveniently expressed in terms of the density matrix $\rho = \rho^\dagger \in \mathbb{C}^{N\times N}$, which describes the probability distribution of a statistical ensemble of quantum states~\cite{breuer2002theory}. One of the most widely used models of these dynamics is the Lindblad master equation~\cite{lindblad1976generators}, which derives from the Markov process generated by a quantum dynamical semigroup.
Specifically, the Lindblad equation for an $N$-dimensional quantum system describes the evolution of $\rho = \rho(t)$ in terms of a Hermitian, time-dependent Hamiltonian $H(t)$ and a set of Lindblad operators $\{L_k\} \subset \mathbb{C}^{N\times N}$, 
\begin{equation}\label{eq:lindblad}
    \dot{\rho} = -\frac{\im}{\hbar} [H, \rho] + \sum_{k=1}^{d}\gamma_k \left(L_k\rho L_k^\dagger - \frac{1}{2}\left\{L_k^\dagger L_k, \rho \right\} \right).
\end{equation}
Here, $\hbar$ denotes the reduced Planck constant, while $[A,B] \equiv AB - BA$ and $\{A,B\} \equiv AB + BA$ denote the commutator and anti-commutator, respectively.
The term $-\im [H, \rho]/\hbar$ in~\eqref{eq:lindblad} describes the unitary evolution of a closed quantum system, equivalent to the Liouville-von Neumann equation, while the summation that follows captures the dissipative effects of the environment on the quantum system.
In the most general setting, there are $d = N^2 - 1$ linearly independent terms in the summation, incorporating a separate relaxation rate $\gamma_k$ for each so-called Lindblad operator $L_k$~\cite{breuer2002theory}. 

Despite being a popular model, the failure of~\eqref{eq:lindblad} to describe thermodynamically-consistent dynamics was first shown in~\cite{grabert1982nonlinear}. Alternative derivations of master equations in Lindblad form have been proposed to achieve more thermodynamically consistent dynamics~\cite{potts2021thermodynamically, trushechkin2021unified}. However, these models only guarantee non-negative entropy production to the same order of the truncation error incurred by the approximations that are necessary for linear master equations. In addition, these models violate the important fluctuation-dissipation theorem~\cite{callen1951irreversibility, kubo1966fluctuation}. Indeed, it is shown in~\cite{grabert1982nonlinear, ottinger2010nonlinear} that \textit{all} linear dissipation models violate the fluctuation-dissipation theorem.

An alternative, \emph{nonlinear} dissipative thermodynamic master equation for open quantum systems was proposed and studied in~\cite{ottinger2010nonlinear, mielke2013dissipative}.
When an isothermal heat bath is used to represent the surrounding environment, this master equation takes the form
\begin{align}\label{eq:quantum_master}
\begin{split}
    \dot{\rho} = -\frac{\im}{\hbar} [H, \rho] &- \sum_{k=1}^d\Big[Q_k, [Q_k, \rho] \Big]\\
    &-\frac{1}{k_B T_e}\sum_{k=1}^d\Big[Q_k, \mathcal{C}_{\rho}[Q_k, H] \Big]
    \,.
    \end{split}
\end{align}    
In this case, the dynamics of $\rho$ are governed by the Hamiltonian $H$ and a set of coupling operators $Q_k = Q_k^\dagger \in \mathbb{C}^{N\times N}$. The temperature of the heat bath is denoted by $T_e$, which we assume to be constant (isothermal). Finally, the super-operator $\mathcal{C}_{\rho}$, known as the \textit{canonical correlation operator}~\cite{mielke2013dissipative}, is defined as
\begin{equation}\label{eq:C_rho}
    \mathcal{C}_{\rho}A \equiv \int_0^1 \rho^s A\rho^{1-s}\dif s
    \,.
\end{equation}
The master equation~\eqref{eq:quantum_master} generalizes to arbitrary thermodynamic environments; see~\cite{ottinger2010nonlinear, ottinger2011geometry} for details.

As shown in~\cite{grabert1982nonlinear, ottinger2010nonlinear}, the nonlinear master equation in~\eqref{eq:quantum_master} is consistent with the fluctuation-dissipation theorem and thus admits thermodynamically admissible dynamics. This is in contrast to the linear Lindblad evolution equation~\eqref{eq:lindblad}. In fact,~\cite{mielke2013dissipative} demonstrates how the nonlinear master equation can be represented in a general framework for nonequilibrium thermodynamics, which we describe next.

\subsection{Thermodynamic consistency}\label{sec:GENERIC}

The GENERIC formalism (General Equation for the Non-Equilibrium Reversible-Irreversible Coupling)~\cite{grmela1997dynamics, ottinger1997dynamics, ottinger2005beyond} is a framework describing the time evolution of nonequilibrium thermodynamic systems as a sum of reversible and irreversible contributions. Modeling a particular system within the context of GENERIC requires the identification of appropriate energy and entropy functionals, along with (state-dependent) linear operators acting on these functionals. 
In~\cite{mielke2013dissipative}, it is shown that~\eqref{eq:quantum_master} is consistent with the GENERIC formalism, which we now describe.

The GENERIC formulation of a dynamical system described by variables $\vx$ takes the form
\begin{equation}\label{eq:generic}
    \dot{\vx} = \mathscr{L}(\vx)\nabla_{\vx}\mathscr{E} + \mathscr{M}(\vx)\nabla_{\vx}\mathscr{S}
    \,,
\end{equation}
where $\mathscr{E}$ and $\mathscr{S}$ denote the energy and entropy of the system, respectively, $\mathscr{L}$ is a skew-symmetric matrix, and $\mathscr{M}$ is a symmetric positive semi-definite matrix. The $\mathscr{L}(\vx)\nabla_{\vx}\mathscr{E}$ term represents reversible (non-dissipative) dynamics and $\mathscr{M}(\vx)\nabla_{\vx}\mathscr{S}$ represents irreversible (dissipative) dynamics. The non-interaction between these two dynamics is encoded in the following compatibility conditions:
\begin{equation}
    \mathscr{L}(\vx)\nabla_{\vx}\mathscr{S} = \mathscr{M}(\vx)\nabla_{\vx}\mathscr{E} = \mathbf{0}
    \,,
\end{equation}
which ensures conservation of energy under irreversible dynamics and that entropy is left unchanged by reversible processes~\cite{ottinger2005beyond}.

When considering the quantum master equation~\eqref{eq:quantum_master} with an isothermal heat bath, the variables of the problem consist of
\begin{equation}
    \vx = \begin{bmatrix}
        \rho\\[1.5pt]
        H_e
    \end{bmatrix},
\end{equation}
where $H_e$ is the internal energy of the surrounding bath. The corresponding energy and entropy functionals are
\begin{align}\label{eq:generic_functionals}
\begin{split}
    \mathscr{E} &= \textrm{Tr}(\rho H) + H_e,\\ \mathscr{S} &= -k_B\textrm{Tr}(\rho \log(\rho)) + \frac{1}{T_e}H_e.
    \end{split}
\end{align}
The gradients of these functionals are given by
\begin{equation}\label{eq:grad_functionals}
    \nabla_{\vx}\mathscr{E} = \begin{bmatrix}
        H\\[1.5pt]
        1
    \end{bmatrix},\quad \nabla_{\vx}\mathscr{S} = \begin{bmatrix}
        -k_B\log(\rho)\\[1.5pt]
        \frac{1}{T_e}
    \end{bmatrix},
\end{equation}
and the GENERIC equations~\eqref{eq:generic} take the form
\begin{align}\label{eq:generic_quantum}
\begin{split}
    \begin{bmatrix}
        \dot{\rho}\\[1.5pt]
        \dot{H_e}
    \end{bmatrix} &= \begin{bmatrix}
        \mathscr{L}_{11} & 0\\[1.5pt]
        0 & 0
    \end{bmatrix}\begin{bmatrix}
        H\\[1.5pt]
        1
    \end{bmatrix} \\[2pt]
    &+ \begin{bmatrix}
        \mathscr{M}_{11} & \mathscr{M}_{12}\\[2.0pt]
        \mathscr{M}_{12}^T & \mathscr{M}_{22}
    \end{bmatrix}\begin{bmatrix}
        -k_B\log(\rho)\\[1.5pt]
        \frac{1}{T_e}
    \end{bmatrix},
    \end{split}
\end{align}
where $\mathscr{L}_{11} = \mathscr{L}_{11}(\rho)$ and $\mathscr{M}_{ij} = \mathscr{M}_{ij}(\rho, Q_k, H)$ depend on the density matrix, the coupling operators, and the Hamiltonian. In~\eqref{eq:generic_quantum}, the reversible term captures the unitary dynamics, i.e.,  $\mathscr{L}_{11}H = -\mathrm{i}[H, \rho]$, while the irreversible portion accounts for the additional terms in~\eqref{eq:quantum_master}.  In the isothermal case, the energy of the environment $H_e$ does not directly contribute to the dynamics of the quantum system. However, for more general thermodynamic environments, the external degrees of freedom will directly affect the quantum system dynamics and must be accounted for explicitly.

\subsection{Neural network models via GFINNs}
Recent dynamics-learning approaches approximate GENERIC systems~\eqref{eq:generic} while enforcing the key properties which ensure thermodynamic consistency~\cite{zhang2022gfinns, gruber2023reversible, hernandez2023port}. In particular, the GFINNs approach taken in~\cite{zhang2022gfinns} parametrizes each of the building blocks $\mathscr{L}$, $\mathscr{M}$, $\mathscr{E}$, and $\mathscr{S}$ separately by feedforward neural networks. Because of this separation, prior knowledge about any of the components can be inserted directly into the governing dynamical system, reducing the complexity of the learned model.

It is natural to approach learning the unknown terms of~\eqref{eq:generic_quantum} by replacing $\mathscr{M}$ by a neural network while using the known forms of the other components. However, the entropy gradient in~\eqref{eq:generic_quantum} is numerically problematic when dealing with quantum systems in a pure state. While the quantum entropy $-k_B\textrm{Tr}(\rho\log(\rho))$ is well-defined for both pure and mixed states, its gradient features a singularity via the term $-k_B\log(\rho)$.

The $\log(\rho)$ singularity does not appear in~\eqref{eq:quantum_master}; indeed, the right-hand side is well-defined and smooth for all valid quantum states. This result relies on the so-called ``miracle relation''~\cite{mittnenzweig2017entropic}, which establishes the following relationship between the canonical correlation operator $\mathcal{C}_\rho$ and $\log(\rho)$,
\begin{equation}\label{eq:miracle_relation}
    \mathcal{C}_{\rho}[A, \log(\rho)] = [A, \rho]
    \,,
\end{equation}
for any linear operator $A$. The GENERIC analog of this relation amounts to the fact that the singularity is canceled out when applying $\mathscr{M}_{11}$ to $-k_B\log(\rho)$ in~\eqref{eq:generic_quantum}.

For this reason, we avoid the separate treatment of $\nabla_{\vx}\mathscr{S}$ and $\mathscr{M}(\vx)$, and instead develop an appropriate architecture for learning the unknown terms in~\eqref{eq:quantum_master} directly. We choose to parametrize the density matrix $\rho$ through its Bloch representation, representing the density matrix as a vector in $\mathbb{R}^d$. In the next section, we review this parametrization and describe how it is used to represent~\eqref{eq:quantum_master} as a system of real-valued ordinary differential equations. Section \ref{sec:nn_dynamics} then describes how the unknown terms are approximated by learnable parameters. 

\section{Bloch-vector Dynamics of the Nonlinear Thermodynamics Master Equation}\label{sec:blochvector_dynamics}

The density matrix of an $N$-level quantum system can be written in the Bloch-vector representation~\cite{bruning2012parametrizations} as
\begin{equation}
    \rho = \frac{1}{N}I + \frac{1}{2}\sum_{j=1}^{d}v_j\sigma_j \equiv \frac{1}{N}I + \frac{1}{2}\vv\cdot\vsigma,
\end{equation}
where $d = N^2 - 1$, the matrices $\sigma_j \in \mathbb{C}^{N\times N}$ are Hermitian matrices forming a basis for Liouville space, and the real vector $\vv$ with entries $(\vv)_j = v_j = \textrm{Tr}(\sigma_j \rho)$ is the \textit{Bloch} vector of $\rho$. 
Similarly, any self-adjoint $N\times N$ operator $A$ can be written as
\begin{equation}\label{eq:mat_bloch}
    A = \frac{a_0}{N}I + \frac{1}{2}\sum_{j=1}^{d}a_j \sigma_j \equiv \frac{a_0}{N}I + \frac{1}{2}\va\cdot\vsigma
\end{equation}
where the coefficients $a_j$ are given by
    $a_0 = \textrm{Tr}(A)$ and $ a_j = \textrm{Tr}(\sigma_j A)$.
%
The basis matrices $\sigma_j$ are the generators of the special unitary group $\textrm{SU}(N)$ and satisfy the following property,
\begin{equation}\label{eq:basis_prop}
    \textrm{Tr}(\sigma_j) = 0\,,\quad \textrm{Tr}(\sigma_i \sigma_j) = 2\delta_{ij}
    \,.
\end{equation}
While there are a number of ways to explicitly define a basis satisfying~\eqref{eq:basis_prop}, we here consider the generalized Gell--Mann matrices defined by
\begin{align} \label{eq:gellman}
    \begin{split}
        \ket{j}\bra{k} + \ket{k}\bra{j},&\\
        -\im\ket{j}\bra{k} + \im\ket{k}\bra{j},&\\
        \left(\frac{2}{\ell(\ell + 1)}\right)^{1/2}\left(\sum_{j=0}^{\ell-1} \ket{j}\bra{j} - \ell\ket{\ell}\bra{\ell} \right),&
    \end{split}
\end{align}
for $0 \leq j < k \leq N-1$ and $1\leq \ell \leq N-1$. In fact, when $N = 2$, these matrices coincide with the two-level Pauli matrices, and coincide with the Gell--Mann matrices when $N=3$. 

Using the Bloch-vector representation, the equation of motion for the density matrix in~\eqref{eq:quantum_master} can be written as a differential equation of the Bloch components
as follows.
Given the generalized Gell--Mann matrices $\sigma_j$, we define the structure constants
\begin{align}\label{eq:fijk_gijk}
\begin{split}
    f_{ijk} &= -\frac{\im}{4}\textrm{Tr}\left(\sigma_i[\sigma_j, \sigma_k]\right),\\
    g_{ijk} &= -\frac{\im}{4}\textrm{Tr}\left(\sigma_i\{\sigma_j, \sigma_k\}\right).
    \end{split}
\end{align}
The constants $f_{ijk}$ define an anti-symmetric tensor, and the $g_{ijk}$ define a totally symmetric tensor. Note that for $N=2$, the entries $f_{ijk}$ coincide with the Levi--Civita symbol, and $g_{ijk} \equiv 0$. 
For $\va \in \mathbb{R}^d$, define the matrix-valued functions $L(\va)$ and $G(\va)$  with entries
\begin{align}\label{eq:LG_def}
\begin{split}
    \left[L(\va)\right]_{ij} = \sum_{k=1}^d a_k f_{ijk}\,,\\
    \left[G(\va)\right]_{ij} = \sum_{k=1}^d a_k g_{ijk}\,.
    \end{split}
\end{align}
The symmetry properties of the structure constants are inherited by these operators, i.e., $L(\va)$ is skew-symmetric and $G(\va)$ is symmetric for every $\va$. 
%
The matrix $L(\cdot)$ is closely related to the commutator of two operators. In particular, if $A$ and $B$ are Hermitian operators with Bloch vector representations $\va$ and $\vb$, respectively, then
\begin{equation}\label{eq:L_commutator}
    \textrm{Tr}\left(\im [A,B]\sigma_j\right) = \left(L(\va)\vb \right)_j = -\left(L(\vb)\va \right)_j.
\end{equation}
The matrix $G(\cdot)$ is related to the anti-commutator of two Hermitian operators as follows:
\begin{align}\label{eq:G_commutator}
    \begin{split}
        \textrm{Tr}\left(\{A, B\}\sigma_j \right) =\ & \frac{2}{N}\left(\textrm{Tr}(A)(\vb)_j + \textrm{Tr}(B)(\va)_j \right)\\
        &+ \left(G(\va)\vb \right)_j.
    \end{split}
\end{align}
In addition, the equality $G(\vb)\va = G(\va)\vb$ holds, and if $A$ and $B$ are traceless, in which case~\eqref{eq:G_commutator} reduces to
\begin{equation}
    \textrm{Tr}\left(\{A, B\}\sigma_j \right) = \left(G(\va)\vb \right)_j = \left(G(\vb)\va \right)_j.
\end{equation}

Using these matrices, we show an explicit representation of the nonlinear thermodynamic master equation~\eqref{eq:quantum_master} for the Bloch vector $\vv$.
Let $\vh$ and $\vq_k$ denote the Bloch vector representation of the Hamiltonian $H$ and the coupling operators $Q_k$, respectively. The Bloch vector representation of the first two terms on the right-hand side of~\eqref{eq:quantum_master} is given by
\begin{align}
    \textrm{Tr}\left(-\frac{\im}{\hbar} [H, \rho]\sigma_j\right) &= \frac{1}{\hbar}\left(L(\vv)\vh\right)_j,\\
    \textrm{Tr}\left(-[Q_k, [Q_k, \rho]]\sigma_j\right) &= \left(L^2(\vq_k)\vv\right)_j,
\end{align}
by equation~\eqref{eq:L_commutator}.
The nonlinear term in~\eqref{eq:quantum_master} is handled by decomposing the super-operator in~\eqref{eq:C_rho} as:
\begin{equation}
    \mathcal{C}_{\rho}A = \frac{1}{2}\{A,\rho\} + \frac{1}{2}\mathcal{C}_{\rho}'A
\end{equation}
where
\begin{equation} \label{eq:C_rho_prime}
    \mathcal{C}_{\rho}'A \equiv -\int_0^1[\rho^s, [\rho^{1-s}, A]]\dif s
\end{equation}
has a double commutator structure.
The Bloch representation of the nonlinear term then has a similar decomposition:
\begin{equation}
    \textrm{Tr}\left(-[Q_k, \mathcal{C}_{\rho}[Q_k,H]]\sigma_j\right) = \left(C(\vv) + D(\vv)\right)_j.
\end{equation}
The first term, $C(\vv)$, is affine in the Bloch-vector $\vv$:
\begin{equation}
    C(\vv) = \frac{1}{N}L(\vq_k)L(\vq_k)\vh + \frac{1}{2}L(\vq_k)G(\vv)L(\vq_k)\vh,
\end{equation}
while the nonlinear contribution takes the form
\begin{equation}
    D(\vv) = -\frac{1}{2}L(\vq_k)M(\vv)L(\vq_k)\vh.
\end{equation}
The matrix $M(\vv)$ is nonlinear in the Bloch vector $\vv$, and its action on a Bloch vector $\va$ for a hermitian operator $A$ corresponds to \eqref{eq:C_rho_prime}. While an explicit expression for $M(\vv)$ is not available for arbitrary $N$, we describe the properties of $M(\vv)$ in the next section, including its approximation by the neural network.

Altogether, the nonlinear thermodynamic master equation~\eqref{eq:quantum_master} is equivalent to the following differential equation for the Bloch vector $\vv$:
\begin{align}\label{eq:bloch_master}
\begin{split}
    &\dot{\vv} = \frac{1}{\hbar}L(\vv)\vh + \sum_{k}L(\vq_k)L(\vq_k)\vv\\
    &+ \frac{\beta}{2}\sum_{k}L(\vq_k)\left(\frac{2}{N}I + G(\vv) - M(\vv)\right)L(\vq_k)\vh
\end{split}
\end{align}
with $\beta = (k_BT_e)^{-1}$.

We briefly discuss the correspondence between~\eqref{eq:bloch_master} and the GENERIC equations given in~\eqref{eq:generic_quantum}. The gradients of the energy and entropy functionals in~\eqref{eq:grad_functionals} are:
\begin{equation}
    \nabla \mathscr{E} = \begin{bmatrix}
        \frac{1}{2}\vh\\[2.5pt]
        1
    \end{bmatrix},\quad \nabla \mathscr{S} = \begin{bmatrix}
        -\frac{k_B}{2}\vw\\[2.5pt]
        \frac{1}{T_e}
    \end{bmatrix}
\end{equation}
where $\vw$ is the Bloch-representation of the matrix $\log(\rho)$.
The blocks in the skew-symmetric matrix $\mathscr{L}$ and the symmetric positive semi-definite matrix $\mathscr{M}$ in~\eqref{eq:generic_quantum} correspond to
\begin{align}
    \mathscr{L}_{11}(\vv) &= \frac{2}{\hbar}L(\vv)\\
    \begin{split}
    \mathscr{M}_{11}(\vv) &= \\-\frac{1}{k_B}&\sum_{k}
    L(\vq_k)\left(\frac{2}{N}I + G(\vv) - M(\vv)\right)L(\vq_k)
    \end{split}\\
        \mathscr{M}_{12}(\vv) &= \mathscr{M}_{21}^T(\vv) = -\frac{1}{2}\mathscr{M}_{11}(\vv)\vh\\
        \mathscr{M}_{22}(\vv) &= \frac{1}{4}\vh^T\mathscr{M}_{11}(\vv)\vh
\end{align}
Thus, the reversible dynamics in~\eqref{eq:bloch_master} correspond to the term $\dfrac{1}{\hbar}L(\vv)\vh$.
The positive semi-definiteness of $\mathscr{M}$ corresponds to the fact that it is the Hessian of the \textit{dual dissipation potential}; see~\cite{mielke2013dissipative} for details.

As mentioned previously, pure quantum states lead to a singularity in the entropy gradient, i.e., in the vector $\vw$. The ``miracle relation'' (equation~\eqref{eq:miracle_relation}), which removes this singularity, corresponds to the fact that
\begin{equation}
    -\frac{1}{2}\mathscr{M}_{11}(\vv)\vw = L(\vq_k)L(\vq_k)\vv,
\end{equation}
which is why no singular terms appear in~\eqref{eq:bloch_master}.

\section{\label{sec:nn_dynamics}Thermodynamically Consistent Learning of the Nonlinear Master Equation}
We describe how each unknown component of~\eqref{eq:bloch_master} can be approximated in a learning framework. Below, we denote by $\theta$ the collection of learnable parameters in the model.

\subsection{Hamiltonian}
The self-adjoint Hamiltonian $H$ is determined by its Bloch representation $\vh$ and its trace $h_0 = \textrm{Tr}(H)$. Because of the double commutator structure of~\eqref{eq:quantum_master}, however, the value of $h_0$ is immaterial when considering the dynamics of $\rho$. This is also apparent from the differential equation for the Bloch vector~\eqref{eq:bloch_master}, where $h_0$ plays no role. Thus, if the Hamiltonian is unknown, it can be learned only up to a translation by a multiple of the identity.
To approximate $H$, we learn a Bloch representation $\vh_\theta \approx \vh$ where each component of $\vh_\theta$ is a learnable parameter.

\subsection{Coupling Operators}
Similarly to the Hamiltonian, the trace of each coupling operator $Q_k$ in~\eqref{eq:quantum_master} does not affect the dynamical evolution of $\rho$. Hence, the operators $Q_k$ are only determined up to translation by a multiple of the identity operator.
Let $\vx^{(k)}$ denote the Bloch vector representation of a traceless coupling operator $Q_k$.
The sum of double commutators between $\rho$ and the coupling operators can then be expressed as
\begin{equation}\label{eq:coupling_mat}
    \sum_{k} \left[Q_k, [Q_k, \rho] \right] = \sum_{i,j=1}^d \Gamma_{ij}\left[\frac{1}{2}\sigma_i, \left[\frac{1}{2}\sigma_j, \rho\right] \right],
\end{equation}
where $\Gamma$ is a semi-definite matrix defined by
\begin{equation}\label{eq:Gamma_def}
    \Gamma_{ij} \equiv \sum_k x_i^{(k)}x_j^{(k)}.
\end{equation}
A similar expressions holds for the term in~\eqref{eq:quantum_master} with the nonlinearity in $\rho$:
\begin{equation}\label{eq:coupling_mat_nonlin}
\sum_{k} \left[Q_k, \mathcal{C}_{\rho}[Q_k,H]\right] = \sum_{i,j=1}^d \Gamma_{ij}\left[\frac{1}{2}\sigma_i, \mathcal{C}_{\rho}\left[\frac{1}{2}\sigma_j, H\right] \right].
\end{equation}
Since the Bloch representation of $\dfrac{1}{2}\sigma_j$ is the canonical basis vector $\ve_j$, then we can express the terms of~\eqref{eq:bloch_master} using the entries of $\Gamma$ and the operators $L_i \equiv L(\ve_i)$. Specifically, letting $\vw \in \mathbb{R}^d$ and $B(\vv) \in \mathbb{R}^{d\times d}$, we have
\begin{align}
    \begin{split}
        \sum_k L^2(\vq_k)\vw &= \sum_{i,j}\Gamma_{ij}L_iL_j\vw,\\
        \sum_k L(\vq_k)B(\vv)L(\vq_k)\vw &= \sum_{i,j}\Gamma_{ij}L_iB(\vv)L_j\vw.
    \end{split}
\end{align}
Equation~\eqref{eq:bloch_master} is therefore equivalent to
\begin{align}\label{eq:bloch_gamma}
    \begin{split}
        &\dot{\vv} = \frac{1}{\hbar}L(\vv)\vh + \sum_{i,j}\Gamma_{ij}L_iL_j\vv\  + \\
         &\frac{\beta}{2}\sum_{i,j}\Gamma_{ij}L_i\left(\frac{2}{N}I + G(\vv) - M(\vv) \right)L_j\vh.
    \end{split}
\end{align}

We are led to the conclusion that learning the coupling operators $Q_k$ is equivalent to learning an approximation $\Gamma_\theta \approx \Gamma$. To enforce semi-definiteness of $\Gamma_\theta$, we parametrize $\Gamma_\theta$ through its Cholesky factorization, $\Gamma_\theta = \widehat{X}_\theta \widehat{X}_\theta^T$ where the entries of the lower-triangular factor $\widehat{X}_\theta$ are learnable parameters. If desired, coupling operators leading to equivalent dynamics can be obtained from the entries $x_j^{(k)} \equiv (\widehat{X}_\theta)_{jk}$ by the expression
\begin{equation}
    Q_k = \frac{1}{2}\sum_{j=1}^{N^2 - 1}x_j^{(k)}\sigma_j.
\end{equation}
We also note that in addition to the entries of $\Gamma$, the temperature of the environment $T_e$ can also be treated as a learnable parameter.

\subsection{Nonlinearity in $\rho$} \label{sec:nonlinearity}
The final piece to reconstructing~\eqref{eq:bloch_master}, or~\eqref{eq:bloch_gamma}, is the approximation of a nonlinear matrix $M(\vv)$ whose action on a Bloch vector $\va$ of a Hermitian operator $A$ corresponds to 
\begin{equation}\label{eq:C_rho_prime}
    \mathcal{C}_{\rho}'A \equiv -\int_0^1[\rho^s, [\rho^{1-s}, A]]\dif s.
\end{equation}
In Appendix A, we derive three necessary conditions on $M(\vv)$ 
for which~\eqref{eq:C_rho_prime} is equivalent to the operation $\va \mapsto -M(\vv)\va$. Specifically, we show that the matrix $M(\vv)$ should:
\begin{enumerate}
    \item be symmetric positive semi-definite;
    \item satisfy the null-space condition $\textrm{null}\left[ L(\vv) \right] \subseteq \textrm{null}\left[M(\vv) \right]$; and
    \item collapse to the operator $L^T(\vv)L(\vv)$ whenever $\vv$ represents a pure quantum state, $\|\vv\|^2 = 2(N-1)/N$.
\end{enumerate}











To approximate $M(\vv)$ and to enforce these three conditions, we take the following approach. We decompose the matrix $M(\vv)$ as follows:
\begin{equation}
    M(\vv) = L^T(\vv)R^T(\vv)R(\vv)L(\vv),
\end{equation}
where $R(\vv)$ is the upper triangular factor of a Cholesky factorization, and $L(\vv)$ is the skew-symmetric matrix defined in~\eqref{eq:LG_def}. The Cholesky factorization ensures positive semi-definiteness, while the multiplication of $L(\vv)$ on the right ensures that the null-space condition is satisfied~\cite{strang2000linear}.

To enforce the third property, construct the matrix $R(\vv)$ as a convex combination of the identity matrix and a triangular matrix $\widetilde{R}(\vv)$, 
\begin{equation}\label{eq:R_decomp}
    R(\vv) = \widetilde{r}^2I + (1 - \widetilde{r}^2)\widetilde{R}(\vv)
\end{equation}
where
\begin{equation} \label{eq:tilde_r}
    \widetilde{r}^2 \equiv \frac{N\|\vv\|^2}{2(N-1)},
\end{equation}
so that pure states are characterized by $\widetilde{r}^2 = 1$. We construct the triangular matrix $\widetilde{R}(\vv)$ by reshaping the output of a feedforward neural network
\begin{equation}\label{eq:NN_R}
    \textrm{NN}_\theta: \mathbb{R}^d \to \mathbb{R}^{\frac{d(d+1)}{2}}
\end{equation}
into the diagonals of an upper triangular matrix.
\subsection{Learning}
The right-hand side of~\eqref{eq:bloch_master} is replaced a parametrized family of functions,
%
\begin{align}\label{eq:F_theta}
\begin{split}
    &\mathcal{F}_{\theta}(\vw) \equiv \frac{1}{\hbar}L(\vw) \vh_\theta + \sum_{i,j}(\Gamma_\theta)_{ij}L_iL_j\vw\\
    &+\frac{\beta}{2}\sum_{i,j}(\Gamma_\theta)_{ij}L_i\left(\frac{2}{N}I + G(\vw) - M_\theta(\vw) \right)L_j\vh_\theta,
    \end{split}
\end{align}
with the corresponding parameterized solution
    $\dot{\vv}_\theta = \mathcal{F}_{\theta}(\vv_\theta)$.
Optimal parameters are found by minimizing the squared $\ell_2$-norm between the predicted and ground-truth Bloch vectors over a set of discrete time values and initial conditions:
\begin{align}\label{eq:loss_single}
\begin{split}
    &\min_\theta \quad \mathcal{L}(\theta) = \\
    &\frac{1}{N_\textrm{ic}}\sum_{j=1}^{N_\textrm{ic}}\frac{1}{N_T}\sum_{i=1}^{N_T}\|\vv_\theta(t_i;\vxi_j) - \vv(t_i;\vxi_j)\|_2^2,\\
    \end{split}
\end{align}
subject to the constraints
\begin{align}
\begin{split}
    \frac{d}{dt}&\vv_\theta(t;\vxi_j) = \mathcal{F}_{\theta}(\vv_\theta(t;\vxi_j)),\\ 
    &\vv_\theta(0;\vxi_j) = \vxi_j.
    \end{split}
\end{align}
Here, $\vv(t_i;\vx_j)$ is the value of the exact trajectory at time $t_i$ starting from the initial condition $\vxi_j$, and $\vv_\theta(t_i;\vxi_j)$ is the prediction of the model. In Section~\ref{sec:three-level} we average over several trajectories beginning at the same initial condition (the ground state), but driven by control Hamiltonians with varying pulse strength.

The true Bloch vectors can be estimated experimentally through quantum state tomography~\cite{paris2004quantum} at selected time steps $t_j$. For our numerical results, we use the complete Bloch vector to define loss functions as in~\eqref{eq:loss_single}. However, when the number of levels $N$ is large, the exponential data requirement to reconstruct the entire density matrix may be prohibitive. Instead, one could define a loss function involving the expectation value of observables $\langle O_i\rangle = \textrm{Tr}(\rho O_i)$ which defines a function of $\vv$.

\section{\label{sec:numerics}Numerical Results}
We first demonstrate the proposed method on a two-level model problem in Section \ref{sec:two-level} for which an explicit form of the nonlinearity is known. 
In Section~\ref{sec:three-level}, we validate the method on a synthetic qutrit experiment trained on data generated by control Hamiltonians with varying pulse strength. Lastly, in Section~\ref{sec:two-lvl-exp}, we employ the learning method to experimental data taken from LLNL's superconducting Quantum Device and Integration Testbed (QuDIT).

\subsection{\label{sec:two-level} Two-Level System}
To demonstrate our proposed method, we first consider a two-level model problem from~\cite{ottinger2010nonlinear}. In addition to its use in quantum optics, the explicit form of the nonlinearity is known, making it an ideal case for verification of learned dynamics.
%
In this test case, we aim to learn the dynamics governed by the Hamiltonian
\begin{equation}
    H = \frac{\hbar\omega}{2}\sigma_3 = \frac{\hbar\omega}{2}\begin{bmatrix}
        1 & 0\\
        0 & -1
    \end{bmatrix},
\end{equation}
and linear coupling determined by the matrix $\Gamma = \widehat{X}\widehat{X}^T$ with $\widehat{X}_{ik} = x_i^{(k)}$ containing the Bloch vector coefficients of the coupling operators in the Pauli basis. The matrix $\widehat{X}$ is chosen to be
\[
    \widehat{X} = \sqrt{\frac{\gamma_1}{2}}\begin{bmatrix}
    1 & 0 & 0 \\
    0 & 1 & 0 \\
    0 & 0 & 0 \\
    \end{bmatrix}.
\]
This choice of $\widehat{X}$ corresponds to coupling which induces longitudinal relaxation or decay, with relaxation time given by $T_1 = \gamma_1^{-1}$~\cite{nielsen2010quantum}.
To generate trajectories for training and testing, we select the parameter values
$\omega = 1.5,\quad \gamma_1 = 0.0785,\quad k_BT_e = 0.65$
and evolve the ODE~\eqref{eq:bloch_gamma} using an explicit formulation of the model starting from 200 initial conditions uniformly sampled from the Bloch sphere. Each trajectory is evolved until $T = 60$, 
and the Bloch vector components are saved every $\Delta t=0.1$. We use 12 of the trajectories to train the model, holding out the remaining 188 for testing.

The Bloch vector of the true Hamiltonian is $\vh/\hbar = (0,\ 0,\ 1.5)^T$, which we approximate with the learnable vector $\vh_\theta$ that is initialized as the zero vector.
We assume no prior knowledge of the coupling operators, and learn the Cholesky factorization of the coupling matrix $\Gamma$ from~\eqref{eq:coupling_mat} starting from a random initialization. In addition, we learn the nonlinear operator $M(\vv)$ following the procedure in Section \ref{sec:nonlinearity}, beginning with randomly initialized layers. 
To train the model, we minimize the loss in~\eqref{eq:loss_single} with $N_\textrm{IC} = 12$ and $N_T = 120$ (corresponding to $T = 12$) using the ADAM optimizer.

After training, we evaluate the model for the initial conditions in the testing set and calculate the average value of the trace distance, $T(\rho_\theta, \rho)$, between the learned density matrix $\rho_\theta$ and the true density matrix $\rho$. For two-level systems, the trace distance can be completely characterized by the $\ell_2$-norm of the corresponding Bloch vectors:
\begin{align}
\begin{split}
    T(\rho_\theta, \rho) &\equiv \frac{1}{2}\textrm{Tr}\left[\sqrt{(\rho_\theta - \rho)^\dagger (\rho_\theta - \rho)} \right]\\
    &= \frac{1}{2}\|\vv_\theta - \vv\|_2.
    \end{split}
\end{align}

\begin{figure}[t]
  \centering
  \includegraphics[width=0.5\textwidth]{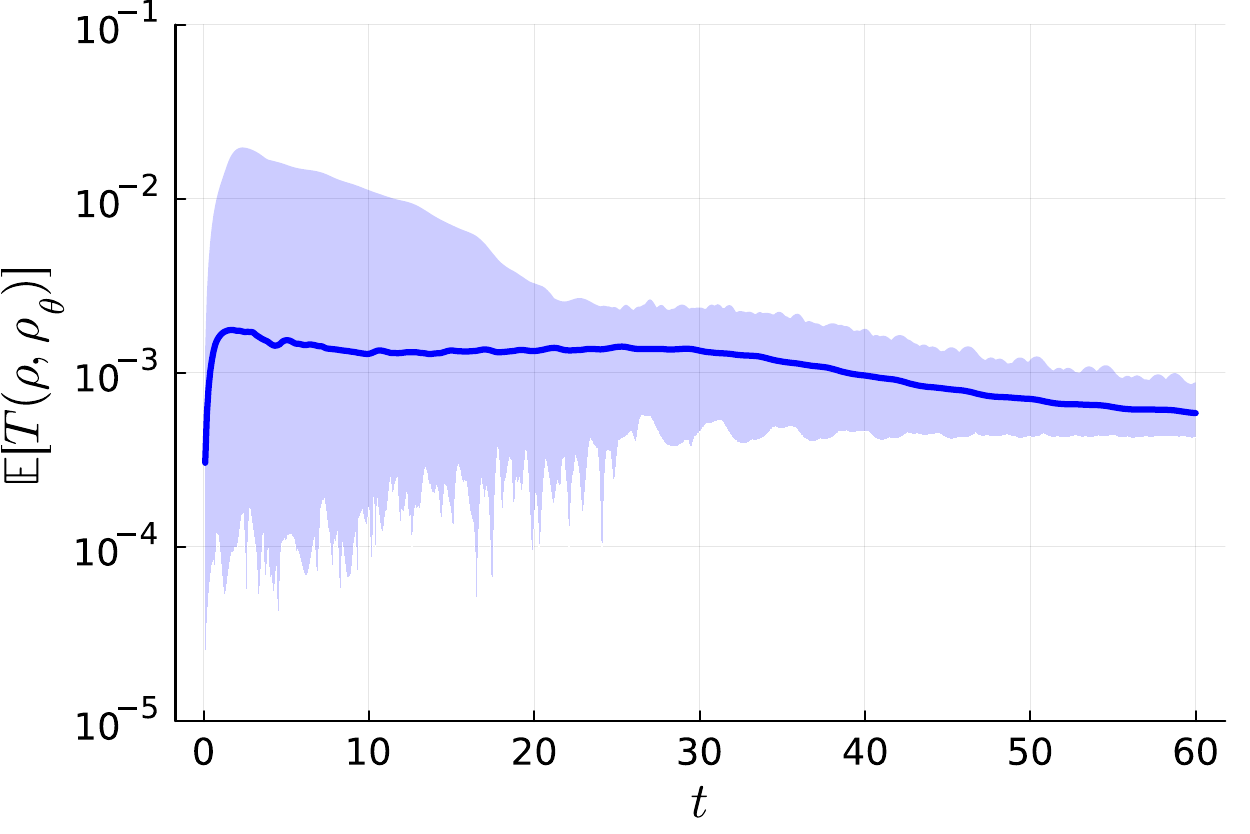}
  \caption{Expectation of trace distance over test trajectories. The shaded region indicates the maximum and minimum values of the trace distance for each time step.}
  \label{fig:two_level_synthetic_trace_dist}
\end{figure}

\begin{figure}
     \centering
     \begin{subfigure}[b]{0.5\textwidth}
         \centering
         \includegraphics[width=\textwidth]{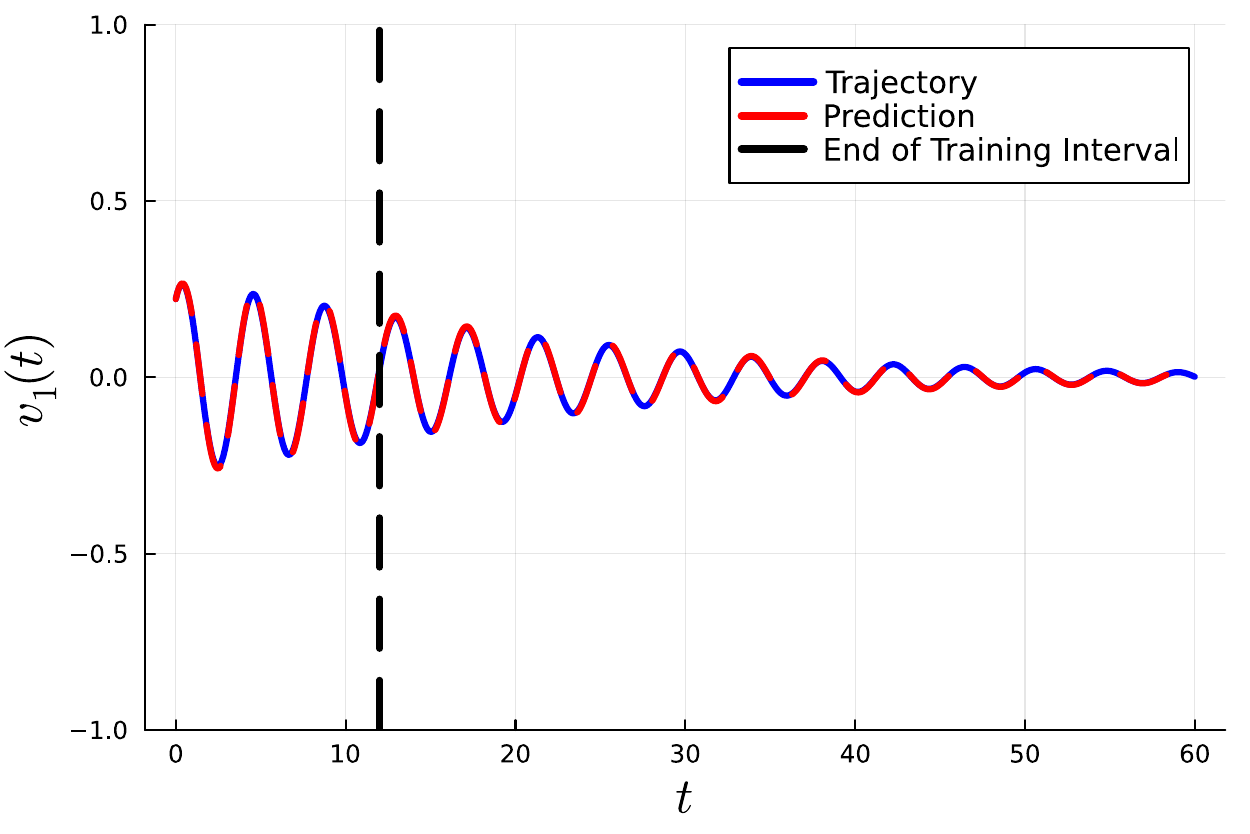}
     \end{subfigure}
     \begin{subfigure}[b]{0.5\textwidth}
         \centering
         \includegraphics[width=\textwidth]{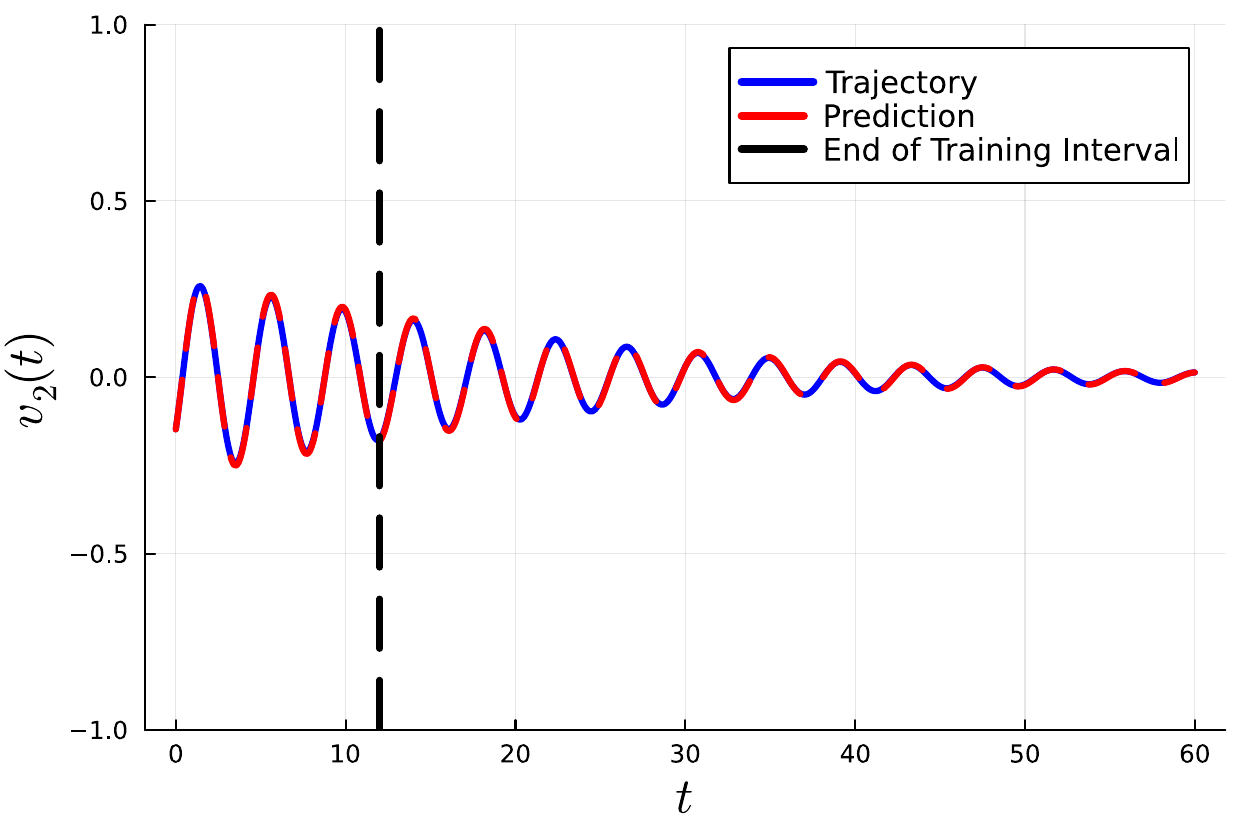}
     \end{subfigure}
     \begin{subfigure}[b]{0.5\textwidth}
         \centering
         \includegraphics[width=\textwidth]{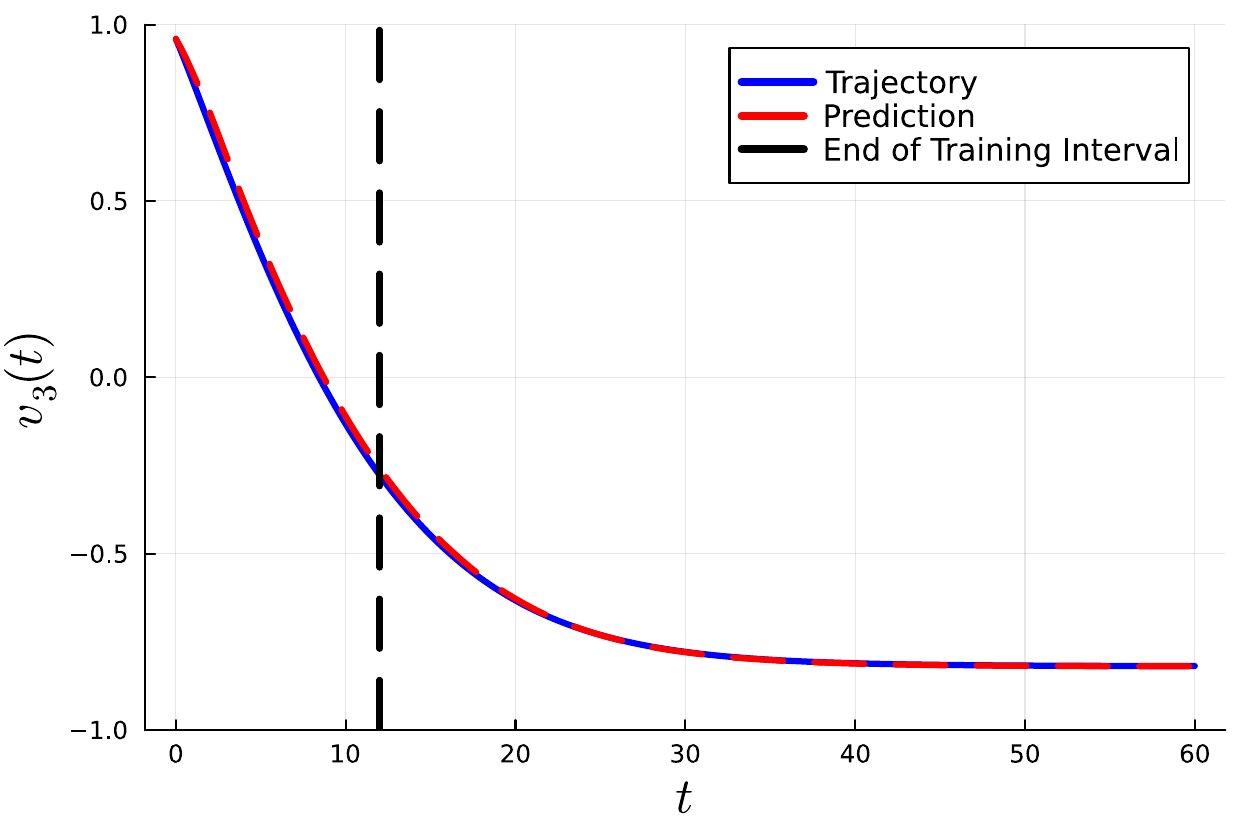}
     \end{subfigure}
        \caption{Predicted Bloch vector trajectories for a two-level quantum system governed by the nonlinear thermodynamic master equation. The trajectory yielding the largest average trace distance over time is chosen. The corresponding initial condition is $v_1(0) = 0.2216$, $v_2(0) = -0.1476$, and $v_3(0) = 0.9596$
        }
        \label{fig:two_level_one_trajectory}
\end{figure}

The average trace distance over the initial conditions for each time step until $T = 60$ is shown in Figure~\ref{fig:two_level_synthetic_trace_dist}. The dynamics of the learned model are compared with the ground truth solution in Figure~\ref{fig:two_level_one_trajectory}. We display the trajectory of the three Bloch vector components corresponding to the initial condition with the largest average trace distance over time. Both the reference solution and the prediction of the model are shown, while the vertical line indicates the end of the training window.

The reconstruction of the system Hamiltonian and the coupling operators is also of interest. The relative error in the learned operators is
\begin{align}
\begin{split}
    \frac{\|\vh_\theta - \vh\|_2}{\|\vh\|_2} &= 6.174\times 10^{-4}\\[2.5pt]
    \frac{\|\widehat{X}_\theta - \widehat{X}\|_F}{\|\widehat{X}\|_F} &= 5.874\times 10^{-2}
    \end{split}
\end{align}
The Hamiltonian is particularly well-resolved, and the learned coupling operators also provide a good approximation to the operators used to generate the synthetic data.

To assess the accuracy further, we compute the error of the nonlinear term in~\eqref{eq:bloch_gamma} using the neural network and learned operators. Specifically, letting
\begin{align}
\begin{split}
    G(\vw) &\equiv \sum_{i,j}\Gamma_{ij}L_iM(\vw)L_j\vh,\\
    G_\theta(\vw) &\equiv \sum_{i,j}(\Gamma_\theta)_{ij}L_iM_\theta(\vw)L_j\vh_\theta,
    \end{split}
\end{align}
we compute the relative error
\begin{align}
\begin{split}
    &\sqrt{\frac{\sum_{j,i} \big\|G_\theta\big[\vv(t_i;\vxi_j) \big] -G\big[\vv(t_i;\vxi_j) \big]\big\|_2^2 }{\sum_{j,i} \big\|G\big[\vv(t_i;\vxi_j) \big]\big\|_2^2}}\\[3pt]
    &= \ 9.727\times 10^{-2},
    \end{split}
\end{align}
where the sum is taken over the 188 initial conditions in the testing set $\vx_j$, and the 600 discrete times $t_i$.

In addition to accuracy, the model also provides interpretability. The learned Bloch vector $\vh_{\theta}$, rounded to two significant digits, is
\begin{equation*}
    \vh_\theta = \begin{bmatrix}
        1.3\times 10^{-4}\\
        6.0\times 10^{-4}\\
        1.5
    \end{bmatrix},
\end{equation*}
while the coupling matrix $\Gamma_\theta = \widehat{X}_\theta\widehat{X}_\theta^T$ is
\begin{equation*}
    \Gamma_\theta = \frac{\gamma_1}{2}\begin{bmatrix}
        1.1 & -6.0\times 10^{-2} & 5.0\times 10^{-4} \\
        -6.0\times 10^{-2}  & 9.4\times 10^{-1} & 2.6\times 10^{-3} \\
        5.0\times 10^{-4}  & 2.6\times 10^{-3} & 1.8\times 10^{-3}
    \end{bmatrix}.
\end{equation*}

From these numerical values, it is clear that the third component of $\vh_\theta$ is the dominant factor of the Hamiltonian part of the ODE, i.e., the unitary dynamics are generated by a diagonal matrix $H$. The entries in the third row/column of $\Gamma_\theta$ indicate that coupling involving the third Pauli matrix $\sigma_3$, dephasing, does not have a strong effect on the dynamics. In contrast, the dominant coupling comes from the first two Pauli matrices $\sigma_1$ and $\sigma_2$, which correspond to decay. Recall that the dissipative dynamics only involve decay, with the dephasing rate set to zero.

These conclusions agree well with the operators used to generate the data. The interpretability of the model may be used to further refine the model, e.g., the learned components with small magnitude can be fixed to zero, and additional training used to refine the approximation of the remaining components.

\begin{figure}
     \centering
     \begin{subfigure}[b]{0.5\textwidth}
         \centering
         \includegraphics[width=\textwidth]{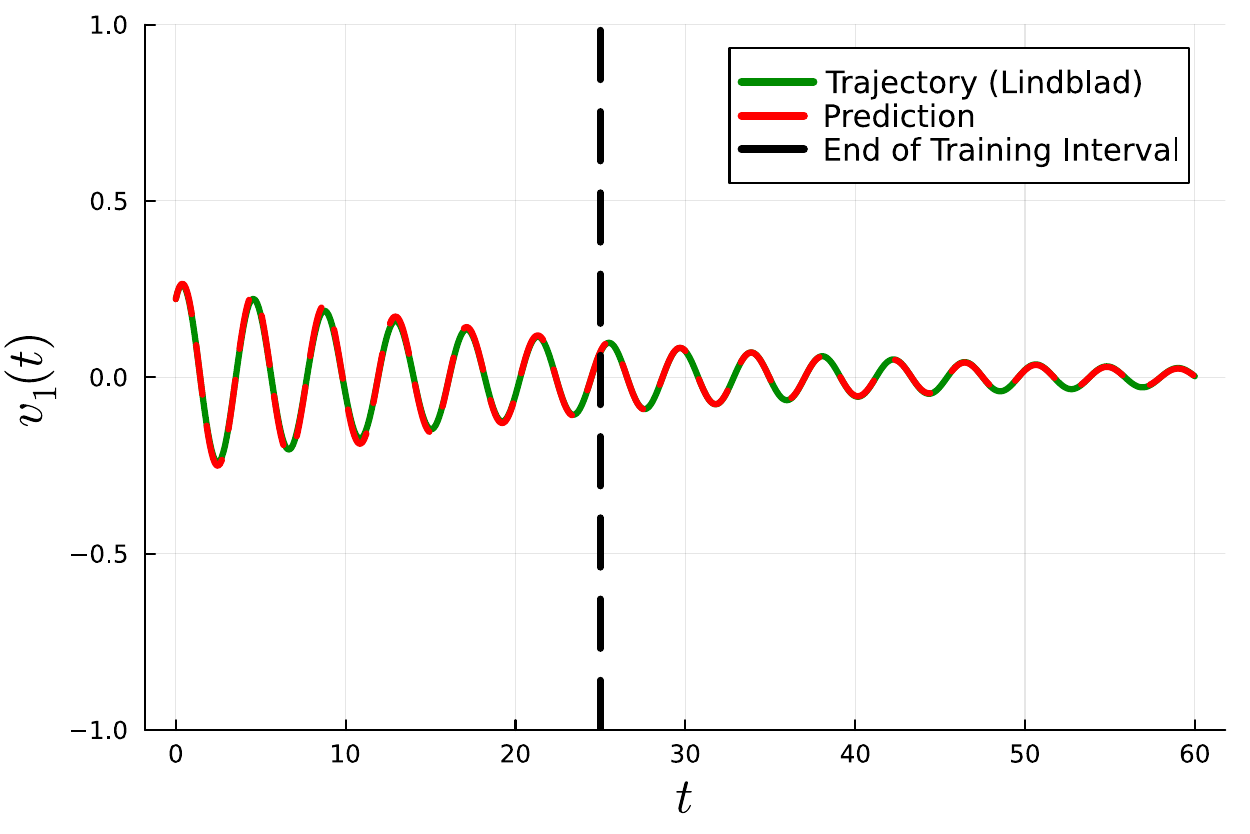}
     \end{subfigure}
     \begin{subfigure}[b]{0.5\textwidth}
         \centering
         \includegraphics[width=\textwidth]{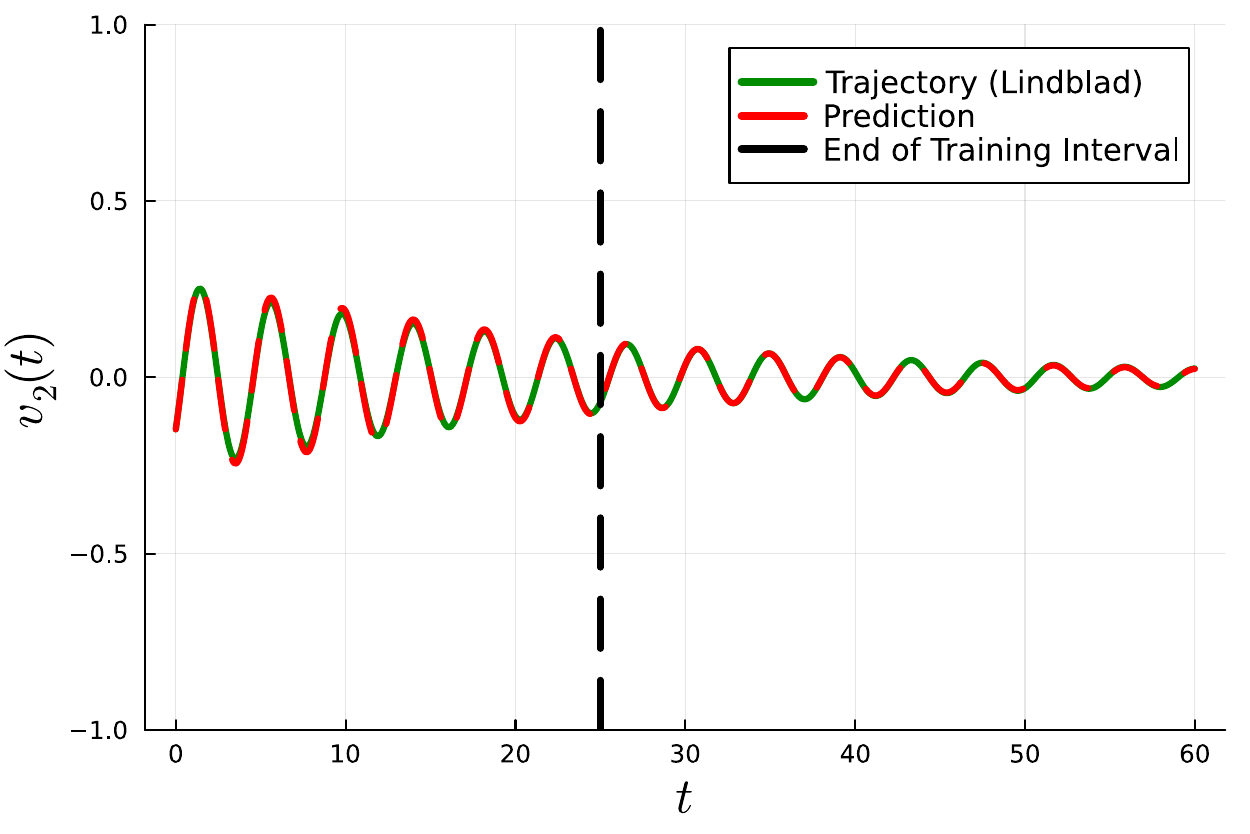}
     \end{subfigure}
     \begin{subfigure}[b]{0.5\textwidth}
         \centering
         \includegraphics[width=\textwidth]{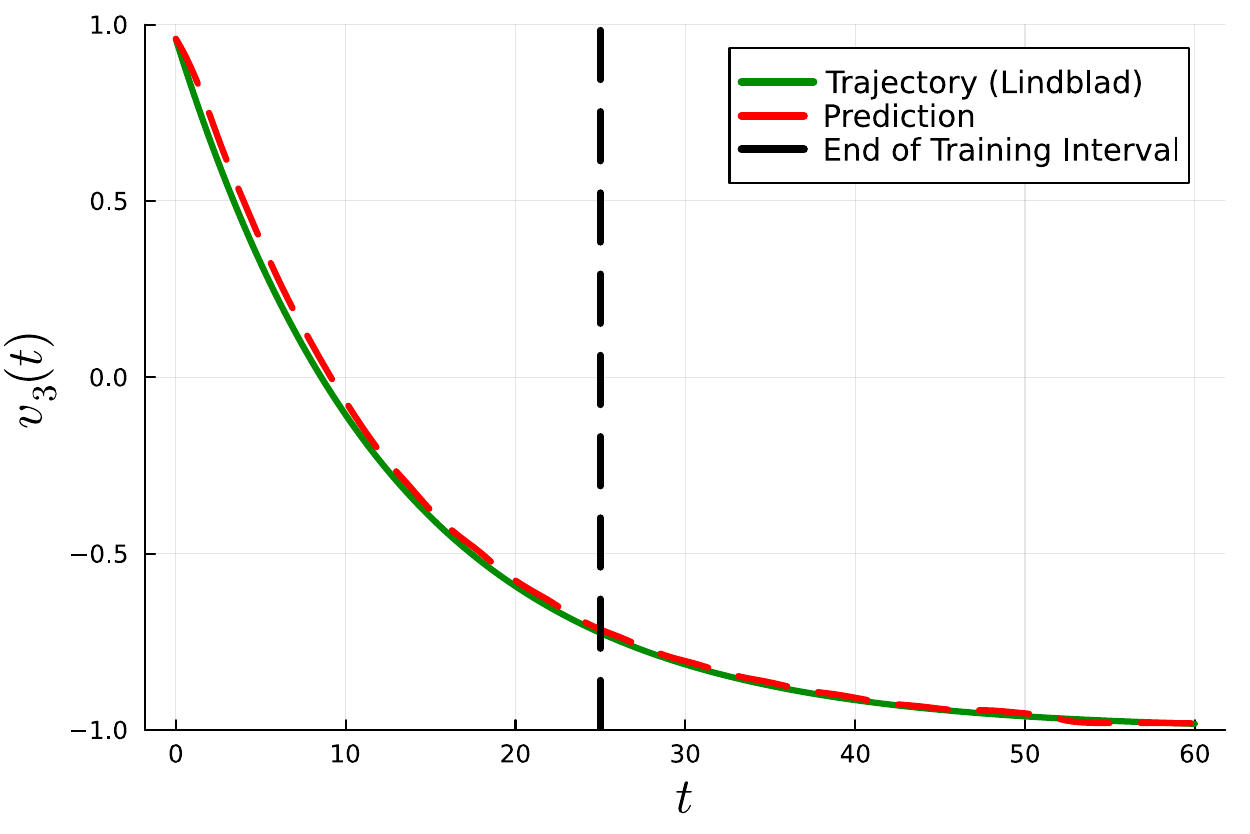}
     \end{subfigure}
        \caption{Predicted Bloch vector trajectories for a two-level quantum system governed by the linear Lindblad equation. The initial condition is identical to the trajectory in Figure~\ref{fig:two_level_one_trajectory}, i.e., $v_1(0) = 0.2216$, $v_2(0) = -0.1476$, and $v_3(0) = 0.9596$.
        }
        \label{fig:two_level_lindblad}
\end{figure}

The classical Lindblad equation is obtained from the nonlinear master equation by (a) linearizing around the thermodynamic equilibrium and (b) letting the temperature go to zero~\cite{mielke2013dissipative}. We illustrate this connection with the synthetic two-level system and demonstrate that the nonlinear learnable model can recover linear Lindblad dynamics as well. 
To do so, we generate training trajectories for the Lindblad equation with parameter values $\omega = 1.5$ and $\gamma_1 = 0.0785$ using the same initial conditions, final time, and timestep size as for the nonlinear trajectories. Figure~\ref{fig:two_level_lindblad} displays the learned model and the ground truth Lindblad solution corresponding to the same initial condition as the trajectory in Figure~\ref{fig:two_level_one_trajectory}. 
While a larger training window was required to achieve comparable accuracy across initial conditions, increasing the final time from $T = 12$ to $T = 25$, this demonstrates that the learned nonlinear model can still be applied to contexts where the dynamics are accurately described by Lindblad equations.

\begin{figure}[t]
  \centering
  \includegraphics[width=0.5\textwidth]{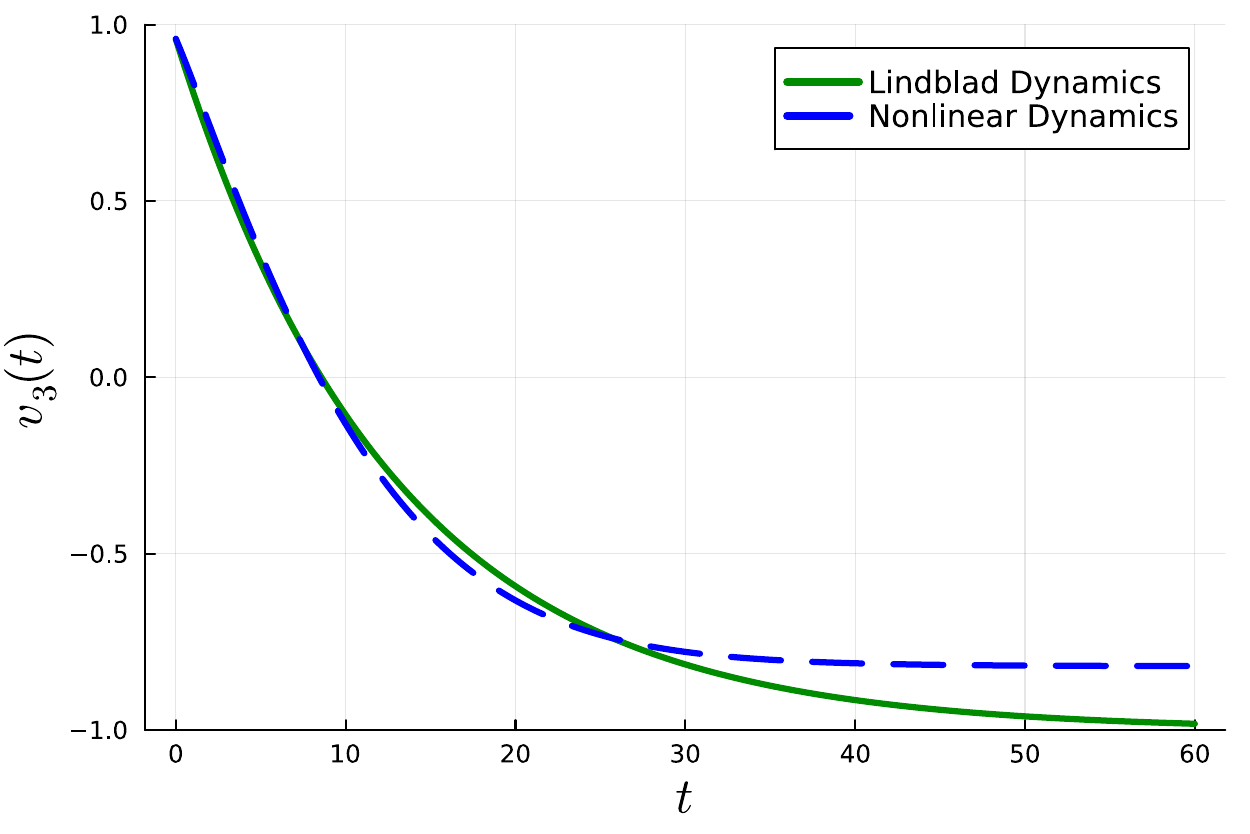}
  \caption{The third Bloch component of the exact trajectories for the Lindblad equation and nonlinear thermodynamic master equation. The initial condition for the Bloch vector is $v_1(0) = 0.2216$, $v_2(0) = -0.1476$, and $v_3(0) = 0.9596$.}
  \label{fig:lindblad_compare}
\end{figure}

As a final comparison between the nonlinear master equation and Lindblad, we plot the third Bloch component for the exact Lindblad solution and exact nonlinear trajectory in Figure~\ref{fig:lindblad_compare}. The nonlinear trajectory approaches equilibrium more rapidly than the solution to the Lindblad equation. In addition, the affect of the zero-temperature limit is visible from the final equilibrium values. The third Bloch component of the Lindblad trajectory approaches $-1$, corresponding to a pure ground state. The nonlinear trajectory, however, converges to a nearby mixed state within the interior of the Bloch sphere.

\subsection{\label{sec:three-level} Synthetic Qutrit System}

For this test case, we consider a Hamiltonian model that describes a superconducting 3-level qutrit that is driven by a microwave control pulse:
\begin{equation} \label{eq:hamiltonian}
    H(t) := \omega a^\dagger a - \xi a^\dagger a^\dagger a a  + f(t) \left(a + a^\dagger \right), 
\end{equation}
where $a$ and $a^\dagger$ are the annihilation and creation operators, respectively, $\omega$ is the qutrit 0-1 transition frequency, $\xi$ is the anharmonicity, and 
\begin{equation}
    f(t)=\Omega(t) \exp(i\omega_d t) + \Omega^*(t) \exp(-i\omega_d t),
\end{equation}
is the control pulse with drive frequency $\omega_d$ and $\Omega(t) = p(t)+iq(t)$. 

To demonstrate our technique, we synthetically generate a data set by solving the nonlinear master equation \eqref{eq:nonlinear-form1} where the nonlinear term is computed analytically (see Appendix C), choosing fixed qutrit frequencies $\omega=344.8$ MHz and $\xi=3.48$ MHz, and coupling operator mimicking realistic decoherence rates of QPUs at LLNL \cite{peng2023qudit}. Specifically, we set the matrix $ \Gamma = \widehat{X}\widehat{X}^T$, with $\widehat{X}_{ik} = x_i^{(k)}$ containing the coefficients of the coupling operators in the Gell-Mann basis (see~\eqref{eq:gellman} and~\eqref{eq:Gamma_def}). The factor $\widehat{X}$ is a block matrix defined as
\begin{equation}
\widehat{X} = \begin{bmatrix}
    \widehat{X}_0\\
    \widehat{X}_1
    \end{bmatrix},
\end{equation}
with
\begin{align}
    \begin{split}
        \widehat{X}_0 &= \begin{bmatrix}
    0.044 & 0 & 0 \\   
    0 & 0.044 & 0 \\
    0 & 0 & -0.16 \\ 
    0 & 0 & 0 
    \end{bmatrix},\\[5pt]
    \widehat{X}_1 &= \begin{bmatrix}
    0 & 0 & 0 \\
    0.07 & 0 & 0 \\  
    0 & 0.07 & 0 \\
    0 & 0 & -0.3   
    \end{bmatrix}.
    \end{split}
\end{align}
%
%
%

To enable efficient numerical simulations, we apply the rotating frame transformation and rotating wave approximation (RWA), 
which transform the system and control Hamiltonian with $\widetilde{H} \approx U H U^\dagger + i\dot U U^\dagger$ for the unitary operator $U = \exp(i H_{d} t)$~\cite{krantz2019quantum}.
For the superconducting qutrit system, we choose $H_{d}=\omega_{d} a^\dagger a$, with $\omega_d = \omega$, which yields the rotating-frame Hamiltonian

\begin{align} \label{eq:rotating_frame_sys}
    \widetilde{H}(t) &= (\omega-\omega_{d}) a^\dagger a - \xi a^\dagger a^\dagger a a + \Omega(t) a + \Omega^*(t) a^\dagger.
\end{align}
Upon application of the rotating frame transformation, the 0-1 transition frequency $\omega$ is shifted to the effective frequency $\omega - \omega_d$. Note that this shift only occurs in the unitary part of the dynamics, while the Hamiltonian in the nonlinear term still involves the original frequency $\omega$.

The training data consists of density matrices obtained by evolving the qutrit system from the initial ground state to a final time $T_{\mathrm{train}}$, driven by control pulses that excite all three levels of the qutrit
\begin{align}
\begin{split}
p(t) = \Omega_{01}^{(p)} + \Omega_{12}^{(p)}( \cos(\xi t) + \sin(\xi t)), \\
q(t) = \Omega_{01}^{(q)} + \Omega_{12}^{(q)}( \cos(\xi t) - \sin(\xi t)).
\end{split}
\end{align}
We generate four different training data trajectories, choosing pulse amplitudes $\Omega_{01}^{(x)} \in \{62.5 \cdot 2^n$ kHz, $n=0,\ldots,3\}$, $\Omega_{12}^{(x)} = \Omega_{01}^{(x)}-0.1$, for $x\in \{p,q\}$. 

Given the synthetically generated training data, we then initialize the coupling operators for training by randomly perturbing their true value by a random variable sampled from $\mathcal{N}(0,0.05)$ and the Hamiltonian by perturbing its last Bloch vector component by a Normal random variable from $\mathcal{N}(0,5)$; this mimics the less precise estimation of the anharmonicity compared to the 0-1 transition frequency (e.g., from Ramsey interferometry \cite{Ramsey1950}). We then aim to relearn them by minimizing the loss function averaged over the four trajectories.
After training, we quantify the accuracy of the learned dynamics by computing the trace distance
between the learned $\rho_{\theta}(t;\Omega)$, and the true density matrices, $\rho(t;\Omega)$ from the synthetic training data set. In this case, the trace distance is given by
\begin{align}
    \begin{split}
        T(\rho, \rho_{\theta}) &= \dfrac{1}{2} \textrm{Tr} \left[ \sqrt{(\rho - \rho_{\theta})^{\dagger}(\rho - \rho_{\theta})} \right]\\
        &= \dfrac{1}{2} \sum_{i=1}^r |\lambda_i|,
    \end{split}
\end{align}
%
where $\lambda_i$ are the eigenvalues of $\rho - \rho_{\theta}$ and $r$ is its rank. 
To quantify the generalization of the learned dynamics to control pulses not seen during training, we compute the expectation of the trace distance over $N_\Omega=100$ randomized control pulses, 
%
\begin{align}
    \begin{split}
        &\mathbb{E}_{\Omega}\left[T(\rho(t;\Omega), \rho_{\theta}(t;\Omega))\right] = \\
        &\dfrac{1}{N_{\Omega}} \sum_{i=0}^{N_{\Omega}} T(\rho(t;\Omega^{(i)}), \rho_{\theta}(t;\Omega^{(i)})),
    \end{split}
\end{align}
%
where $\Omega^{(i)} \sim \mathcal{U}(\Omega_L, \Omega_U)$ is the $i$-th sample from a uniform distribution $\mathcal{U}(\Omega_L, \Omega_U)$ with support on $[\Omega_L, \Omega_U]$ and $t \in [0,20]\mu\mathrm{s}$. 

\begin{figure}
    \centering 
    \begin{subfigure}[b]{0.4\textwidth}
        \centering
        \includegraphics[width=\textwidth]{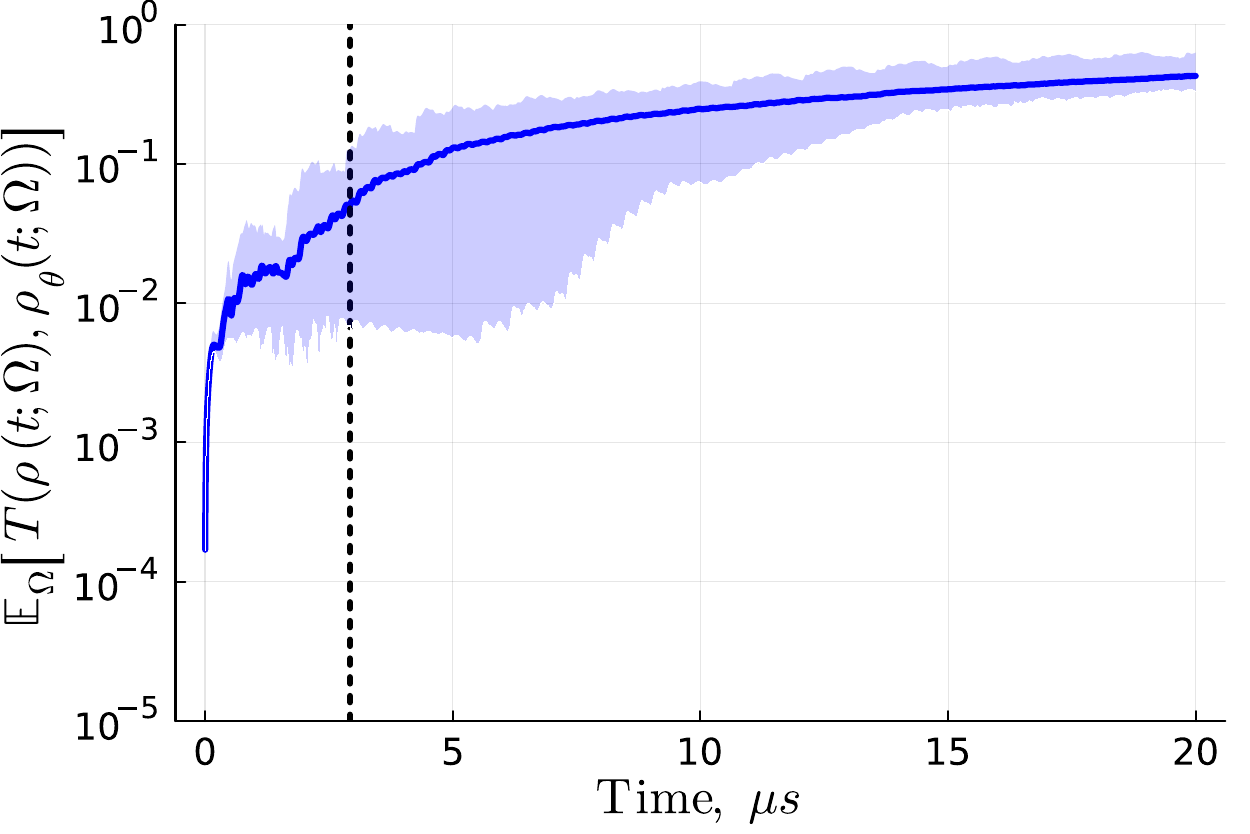}
    \end{subfigure}
    \begin{subfigure}[b]{0.4\textwidth}
        \centering
        \includegraphics[width=\textwidth]{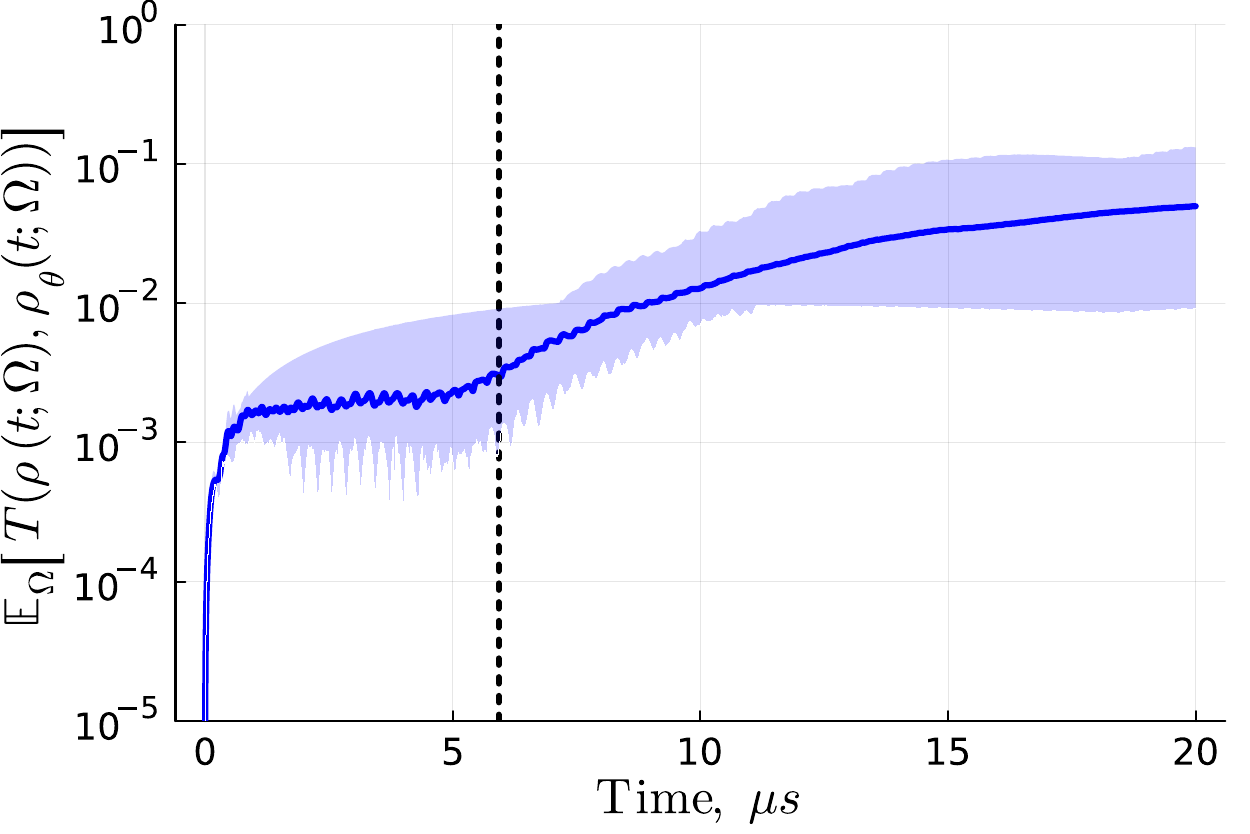}
    \end{subfigure}
    \begin{subfigure}[b]{0.4\textwidth}
        \centering
        \includegraphics[width=\textwidth]{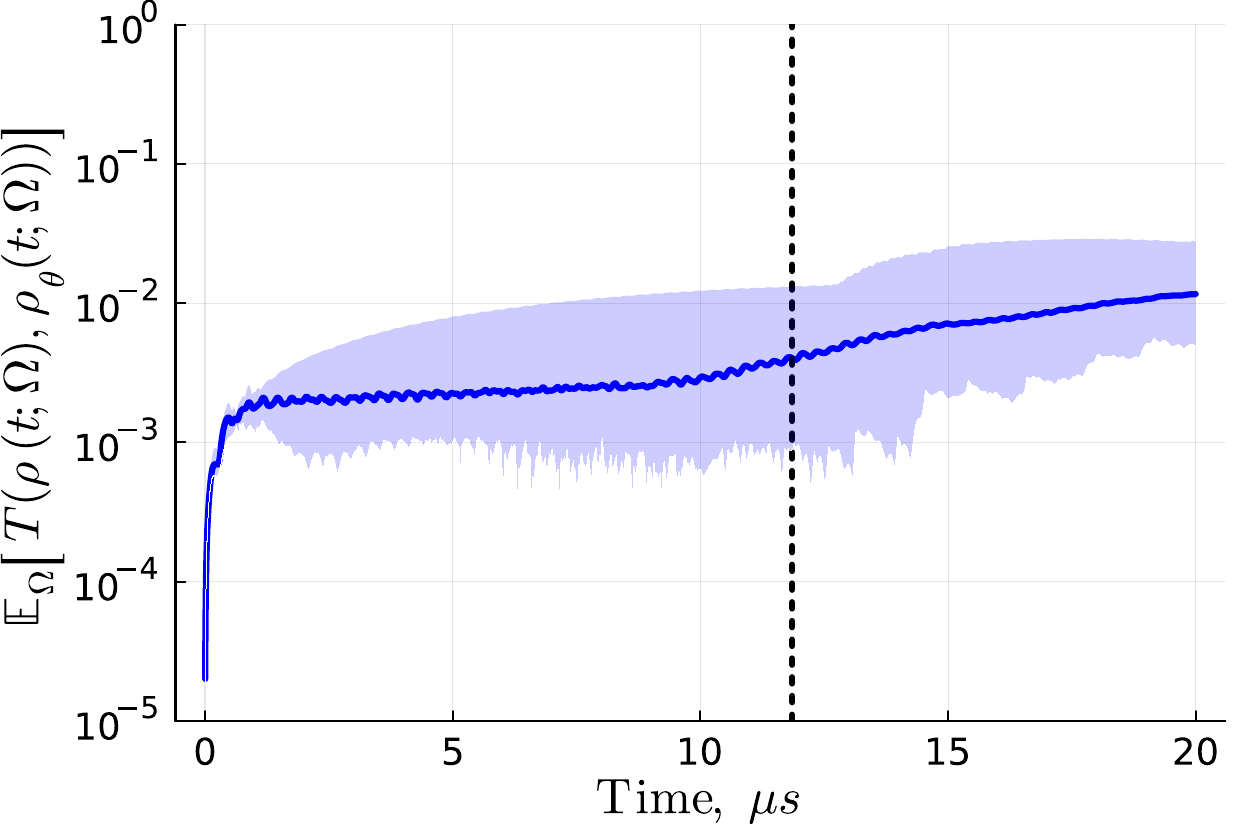}
    \end{subfigure}
       \caption{Expectation (solid line) and maximum and minimum (shaded region), over all controls, of trace distance between the density matrices from learned and true dynamics of a qutrit system trained over different training intervals (black vertical lines).}
       \label{fig:three_level_synthetic_exptrace}
\end{figure}

Figure \ref{fig:three_level_synthetic_exptrace} shows the expectation and the bounds (i.e., maximum and minimum) of the trace distance between the density matrices from the learned and true dynamics of the qutrit system with different training windows. We see that, for sufficiently long training windows, the expected trace distance within the training window is of similar magnitudes for the three intervals analyzed. For the evolution beyond the training window, the expected trace distance increases with time, with the longer training windows yielding more accurate (i.e., lower trace distances and tighter bounds) evolution. For $T_{\mathrm{train}} = 12\mu \mathrm{s}$, the expected trace distance is less than $0.01$ for $t \leq 20\mu \mathrm{s}$ with the maximum trace distance being less than $0.05$. 

\begin{figure}
    \centering 
    \begin{subfigure}[b]{0.4\textwidth}
        \centering
        \includegraphics[width=\textwidth]{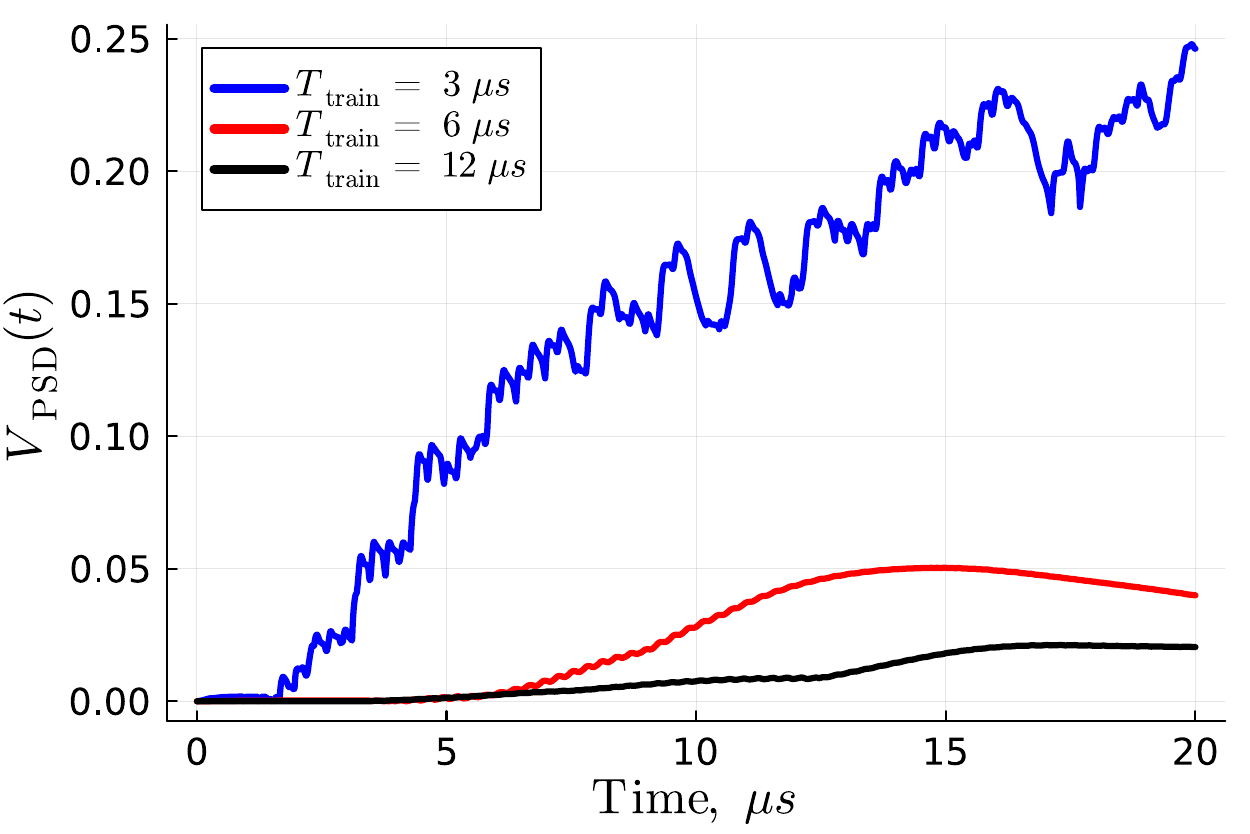}
    \end{subfigure}
       \caption{PSD-violation of the density matrices from learned dynamics of a qutrit system trained over different intervals $T_{\mathrm{train}}$.}
       \label{fig:three_level_synthetic_PSDVio}
\end{figure}

The formulation presented in Section \ref{sec:nonlinearity} restricts the dynamics to a unit sphere, \eqref{eq:tilde_r}, and is sufficient to guarantee physically consistent evolution of a single qubit system.
For general $N$ level systems, however, the set of admissible quantum state is only a subset of the hypersphere 
\cite{bruning2012parametrizations,kimura2003}. 
Nevertheless, we show here that the learned dynamics do in fact yield a physically consistent evolution of the qutrit system, given physically consistent training data.
We quantify the violation of positive semi-definiteness (PSD) of the density matrix, $V_{\mathrm{PSD}}$, by the magnitude of the largest (in absolute value) negative (i.e., smallest) eigenvalue of the density matrices from the learned operators over the controls $\Omega$
\begin{equation}
    V_{\mathrm{PSD}}(t) := \left| \min \left\{ 0,~\min_\Omega \lambda\left(\rho(t;\Omega)\right) \right\} \right|,
\end{equation}
%
where $\lambda\left(\rho\right)$ is the spectrum of $\rho$. Note that $V_{\mathrm{PSD}}(t) \equiv 0$ corresponds to a PSD-preserving map, which, together with $\mathrm{Tr}(\rho(t)) \equiv 1$ is a CPTP map.
Figure \ref{fig:three_level_synthetic_PSDVio} shows the PSD-violation of the density matrices from the learned dynamics trained on different training intervals. We see that for $T_{\mathrm{train}} = 12\mu \mathrm{s}$, $V_{\mathrm{PSD}} \leq 0.02$ within the training window and remains $V_{\mathrm{PSD}} \leq 0.03$ for $t \leq 20\mu \mathrm{s}$. For $T_{\mathrm{train}} = 3\mu \mathrm{s}$, the magnitude of PSD-violation increases drastically with time. For $T_{\mathrm{train}} = 3\mu \mathrm{s} ,~6\mu \mathrm{s},~12\mu \mathrm{s}~$, the fraction of density matrices in the validation data that violated the PSD condition was 84\%, 4\% and 4\%, respectively. We can report that the trace of all density matrices was preserved to machine precision. Hence, for a sufficiently long training window, the proposed method can learn the dynamics of a qutrit system and preserve positive semi-definiteness of the density matrices to high accuracy and preserve the trace to machine precision.

\subsection{\label{sec:two-lvl-exp} Experimental Data}
In this section, we validate our model with experimental data taken at LLNL's QUDIT device, compare Appendix D for details of the experimental setup for data generation. 
In order to improve stability of the model, we performed training on increasingly larger time domains. In total, we trained over 25 increments, each covering 30 timesteps, until a total training domain of 14.98 $\mu\mathrm{s}$ is reached. See Appendix E for details on the neural network hyperparameters.

We found that the Adam optimizer exhibited slightly worse performance when the training time domain was extended by each increment. To mitigate this, we trained the model using the L-BFGS \cite{nocedal2006numerical} optimizer. We emphasize that this choice was made during training, and was not based on performance for the testing set (corresponding to times after 14.98 $\mu\mathrm{s}$).
To improve training convergence, a Tikhonov $\ell^2$ penalty with coefficient $\lambda = 10^{-3}$ was added to the weights of the neural network. The Hamiltonian and coupling parameters were not regularized in any way.

\begin{figure}
    \centering 

    \begin{subfigure}[b]{0.5\textwidth}
        \includegraphics[width=1\textwidth]{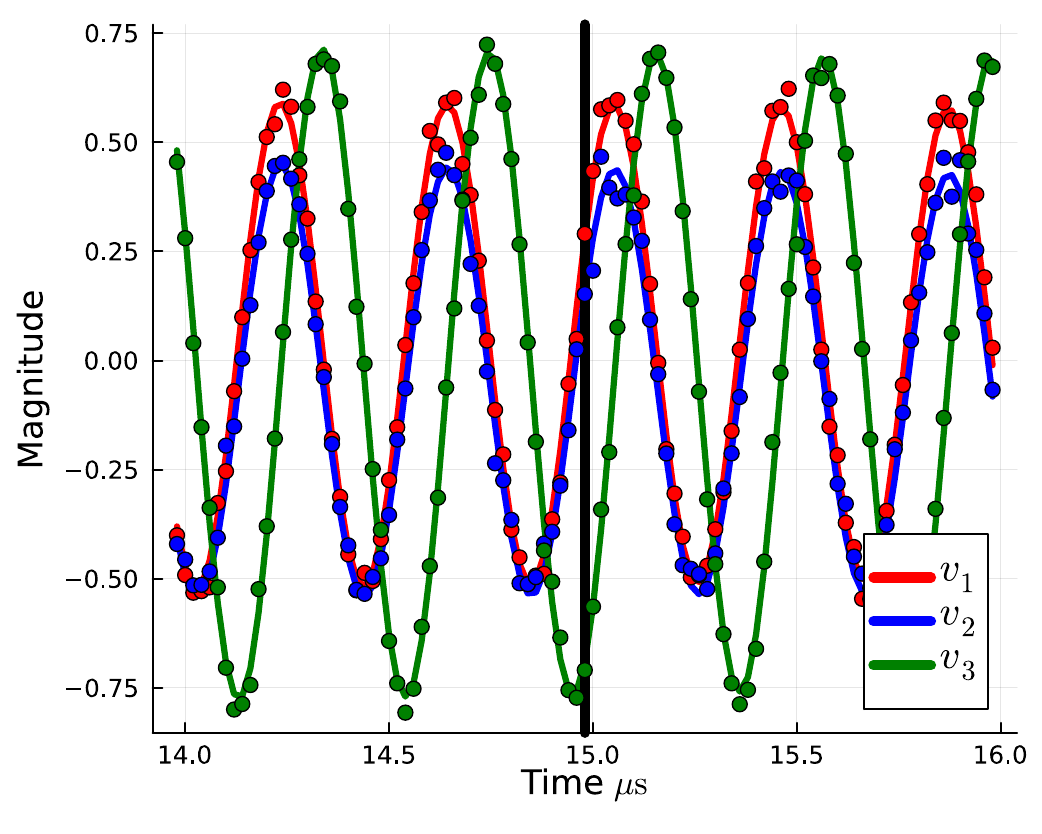}
        \subcaption{The black line indicates the end of the training interval. Extrapolation of learned trajectory immediately after training.}
        \label{fig:data_three_level_trajectories_before_after}
    \end{subfigure}
    \begin{subfigure}[b]{0.5\textwidth}
        \includegraphics[width=1\textwidth]{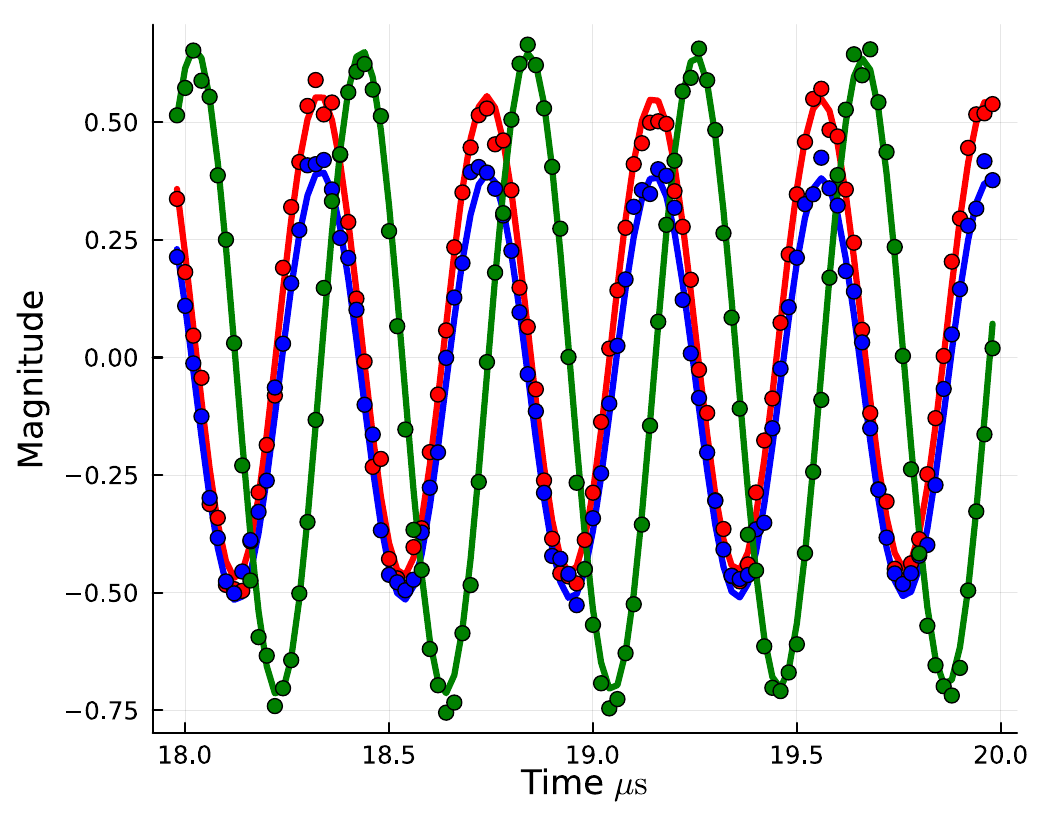}
        \subcaption{Extrapolation 5 microseconds after training.}
        \label{fig:data_three_level_trajectories_far_after}
    \end{subfigure}
    
    \caption{Predicted trajectory after training up to 14.98 $\mu \mathrm{s}$ (lines), compared to data taken from LLNL's QPU (dots). The top figure plots the predicted trajectory $1\mu \mathrm{s}$ before and after the training period. The bottom plot demonstrates the extrapolation to data $5 \mu \mathrm{s}$ past the training domain.}
   \label{fig:data_three_level_trajectories}
\end{figure}

\begin{figure}
    \centering
    \includegraphics[width=0.5\textwidth]{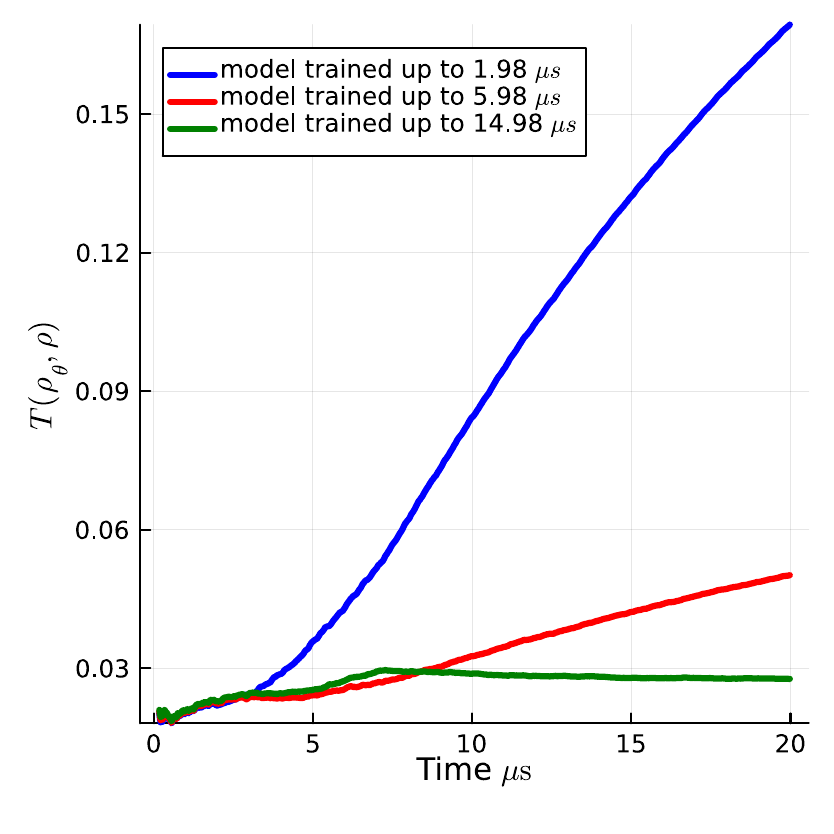}
       \caption{Cumulative trace distances (defined as $\frac{1}{N}\sum_{i=1}^{N} T(\rho_\theta(t_i),\rho(t_i))$ for $N=1,\ldots,N_T$) between the trained model prediction of density matrix $\rho_{\theta}$ and the data ${\rho}$ during training. We see that the cumulative error plateaus when the model was trained to $14.98\mu \mathrm{s}$.}
       \label{fig:cumulative_trace_distances_3level}
\end{figure}

We first train on our model to match a single data trajectory generated with pulse strengths of $p = q= 0.0181 \times 47.9$ MHz. Figure \ref{fig:data_three_level_trajectories}, plots the evolution of the learned model Bloch vector components (lines) alongside the data (dots), right before and after the total training domain of $14.98 \,\mu\mathrm{s}$ (Figure \ref{fig:data_three_level_trajectories_before_after}) and far after (Figure \ref{fig:data_three_level_trajectories_far_after}), demonstrating that the learned model extrapolates well to long-time data it has not seen during training. The cumulative errors over time are quantified in Figure \ref{fig:cumulative_trace_distances_3level}, for three different training domains. Most importantly, we see that the cumulative error remains constant beyond the training domain of $14.98\mu \mathrm{s}$, leveling at $~3\times10^{-2}$. We found comparable performance when the network was trained on data downsampled up to a factor of 100.

This demonstrates that the neural network has learned the underlying dynamics from noisy data and extrapolates outside of the training window.

Interestingly, the learned factor $\widetilde{R}(\vv)$ (in (\ref{eq:R_decomp})) was constant (up to machine precision) with respect to variations in the Bloch vector $\vv$. Therefore, in this dataset, the learned nonlinearity is fully determined by the norm of the Bloch vector. Whether this is true for other devices or other quantum processes offers an interesting future direction for research.

Adding independent Gaussian noise with standard deviation 0.1 to each Bloch component, while clamping data to ensure it is between $-1$ and $1$, at each time step required training with a larger time increment of 50 and increased the final trace distance error roughly 3-fold. 

\begin{figure}
    \centering
    \includegraphics[width=0.5\textwidth]{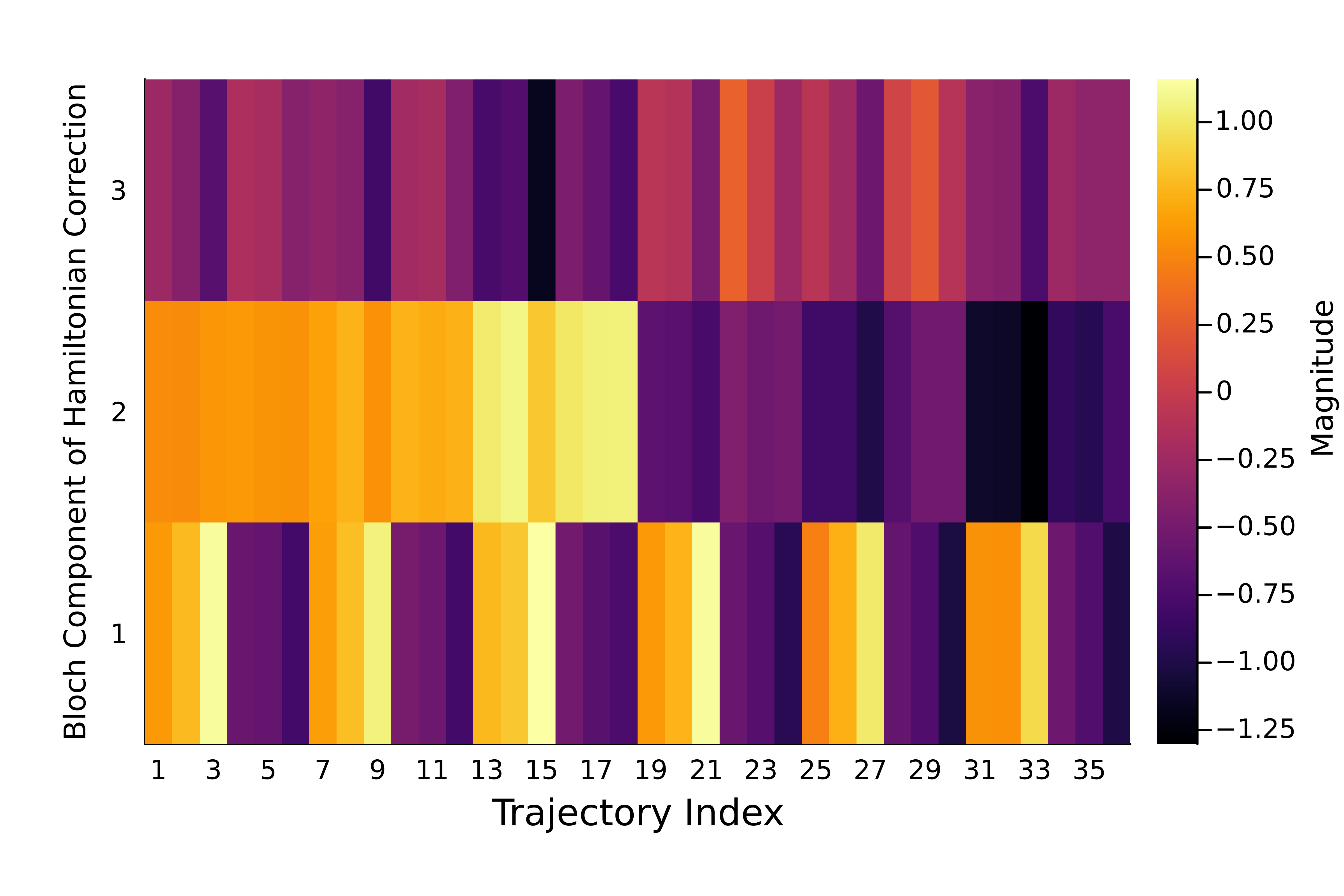}
       \caption{Learned $\vh_\theta$ components for 36 different constant $p$ and $q$ pulse strengths. Each column are the 3 components of $\vh_\theta$ when the neural network is trained on an individual control pulse trajectory. There is a clear pattern across different pulse strengths that motivates future modeling of parameter dependence across individual trajectories.}
       \label{fig:heatmap_learned_hamiltonians}
\end{figure}

Qualitatively similar results were observed when training on single trajectory data with different control pulse amplitudes. 
However, we observed that the trained model parameters strongly depend on the pulse strengths $p, q$, as can be seen in Figure \ref{fig:heatmap_learned_hamiltonians}. Consequently, training a model on multiple trajectories leads to an averaged model that does not fit well to any of the individual trajectories. This observation motivates future work on learning models that incorporate pulse-dependent parameters to capture the effects of pulse distortions. Such a model would allow for more comprehensive training, resulting in a model that extrapolates to different control pulses. 

\section{\label{sec:conclusion}Summary and Conclusion}

In this work, we developed a data-driven framework for characterizing the dynamics of open quantum systems that yield thermodynamically consistent predictions. Our approach employs the GENERIC framework for constructing a structure-preserving, thermodynamically consistent parameterization of the nonlinear master equation governing quantum state evolution. By embedding fundamental physical constraints within the model ansatz, the method ensures physically valid predictions of the evolution of quantum states. We demonstrate our approach on synthetic test problems for characterizing a single qubit as well as a qutrit system, and perform training on experimental data taken from a quantum processing unit at LLNL’s QuDIT. We demonstrate that, for sufficiently long training time intervals, our method preserves positive semi-definiteness of the density matrices to high accuracy and trace of the density matrices to machine precision. Furthermore, the approach was able to accurately predict the dynamics of the quantum system beyond the training window.

The results highlight the capability of the method to learn and generalize quantum dynamical models with high fidelity, offering a robust tool for studying dissipative quantum processes and non-equilibrium dynamics. Furthermore, the learned model can serve as a digital twin of the quantum system, providing a computationally efficient and accurate surrogate that can be leveraged for optimizing control pulses and validating noise mitigation techniques. 

In order to extend the applicability of this framework, future research should explore how feasible computational cost can be achieved in large multi-qubit systems. It is pointed out in~\cite{gruber2023reversible} that the complexity of training standard metriplectic architectures grows rapidly as the problem size increases. In order to handle large, multi-qubit systems with computational efficiency, more sophisticated thermodynamically consistent models may be required. The adaptation of metriplectic graph neural networks~\cite{gruber2023reversible} or port-metriplectic architectures designed to handle interacting thermodynamic subsystems~\cite{hernandez2023port} to multi-qubit systems is a natural direction for future work.
\section*{Acknowledgments}
This work was performed under the auspices of the U.S. Department of Energy by Lawrence Livermore National Laboratory under Contract DE-AC52-07NA27344. LLNL-JRNL-2005344.

Brendan Keith was supported in part by the National Science Foundation under Grant No.\ 2407452.

Peter Sentz’s work is supported by the NSF-DMS 2038039.

This material is based upon work supported by the National Science Foundation Graduate Research Fellowship under Grant No.\ 2439559.

\bibliographystyle{quantum}
\bibliography{refs}

\providecommand{\noopsort}[1]{}\providecommand{\singleletter}[1]{#1}%
\begin{thebibliography}{10}

\bibitem{corcoles2019challenges}
Antonio~D. C{\'o}rcoles, Abhinav Kandala, Ali Javadi-Abhari, Douglas~T.
  McClure, Andrew~W. Cross, Kristan Temme, et~al.
\newblock ``Challenges and opportunities of near-term quantum computing
  systems''.
\newblock \href{https://dx.doi.org/10.1109/JPROC.2019.2954005}{Proceedings of
  the IEEE {\bf 108}, 1338--1352}~(2020).

\bibitem{tuckett2019tailoring}
David~K. Tuckett, Andrew~S. Darmawan, Christopher~T. Chubb, Sergey Bravyi,
  Stephen~D. Bartlett, and Steven~T. Flammia.
\newblock ``Tailoring surface codes for highly biased noise''.
\newblock \href{https://dx.doi.org/10.1103/PhysRevX.9.041031}{Physical Review X
  {\bf 9}, 041031}~(2019).

\bibitem{gebhart2023learning}
Valentin Gebhart, Raffaele Santagati, Antonio~Andrea Gentile, Erik~M. Gauger,
  David Craig, Natalia Ares, Leonardo Banchi, Florian Marquardt, Luca
  Pezz{\`e}, and Cristian Bonato.
\newblock ``Learning quantum systems''.
\newblock \href{https://dx.doi.org/10.1038/s42254-022-00552-1}{Nature Reviews
  Physics {\bf 5}, 141--156}~(2023).

\bibitem{baldwin2014quantum}
Charles~H. Baldwin, Amir Kalev, and Ivan~H. Deutsch.
\newblock ``Quantum process tomography of unitary and near-unitary maps''.
\newblock \href{https://dx.doi.org/10.1103/PhysRevA.90.012110}{Physical Review
  A {\bf 90}, 012110}~(2014).

\bibitem{nielsen2010quantum}
Michael~A. Nielsen and Isaac~L. Chuang.
\newblock ``Quantum computation and quantum information''.
\newblock \href{https://dx.doi.org/10.1017/CBO9780511976667}{Cambridge
  University Press}. ~(2010).
\newblock 10th {A}nniversary edition.

\bibitem{cerrillo2014non}
Javier Cerrillo and Jianshu Cao.
\newblock ``Non-{M}arkovian dynamical maps: {N}umerical processing of open
  quantum trajectories''.
\newblock \href{https://dx.doi.org/10.1103/PhysRevLett.112.110401}{Physical
  Review Letters {\bf 112}, 110401}~(2014).

\bibitem{samach2022lindblad}
Gabriel~O. Samach, Ami Greene, Johannes Borregaard, Matthias Christandl, Joseph
  Barreto, David~K. Kim, Christopher~M. McNally, Alexander Melville, Bethany~M.
  Niedzielski, et~al.
\newblock ``Lindblad tomography of a superconducting quantum processor''.
\newblock \href{https://dx.doi.org/10.1103/PhysRevApplied.18.064056}{Physical
  Review Applied {\bf 18}, 064056}~(2022).

\bibitem{huang2023learning}
Hsin-Yuan Huang, Sitan Chen, and John Preskill.
\newblock ``Learning to predict arbitrary quantum processes''.
\newblock \href{https://dx.doi.org/10.1103/PRXQuantum.4.040337}{PRX Quantum
  {\bf 4}, 040337}~(2023).

\bibitem{hangleiter2024robustly}
Dominik Hangleiter, Ingo Roth, Jon{\'a}{\v{s}} Fuksa, Jens Eisert, and Pedram
  Roushan.
\newblock ``Robustly learning the {H}amiltonian dynamics of a superconducting
  quantum processor''.
\newblock \href{https://dx.doi.org/10.1038/s41467-024-52629-3}{Nature
  Communications {\bf 15}, 9595}~(2024).

\bibitem{mohseni2024deep}
Naeimeh Mohseni, Junheng Shi, Tim Byrnes, and Michael~J. Hartmann.
\newblock ``Deep learning of many-body observables and quantum information
  scrambling''.
\newblock \href{https://dx.doi.org/10.22331/q-2024-07-18-1417}{Quantum {\bf 8},
  1417}~(2024).

\bibitem{zhu2023quantum}
Yan Zhu, Ya-Dong Wu, Qiushi Liu, Yuexuan Wang, and Giulio Chiribella.
\newblock ``Quantum process learning through neural emulation''~(2023).
\newblock  \href{http://arxiv.org/abs/2308.08815}{arXiv:2308.08815}.

\bibitem{lewis2024improved}
Laura Lewis, Hsin-Yuan Huang, Viet~T. Tran, Sebastian Lehner, Richard Kueng,
  and John Preskill.
\newblock ``Improved machine learning algorithm for predicting ground state
  properties''.
\newblock \href{https://dx.doi.org/10.1038/s41467-024-45014-7}{Nature
  Communications {\bf 15}, 895}~(2024).

\bibitem{liu2022solving}
Zidu Liu, L.-M. Duan, and Dong-Ling Deng.
\newblock ``Solving quantum master equations with deep quantum neural
  networks''.
\newblock \href{https://dx.doi.org/10.1103/PhysRevResearch.4.013097}{Physical
  Review Research {\bf 4}, 013097}~(2022).

\bibitem{hartmann2019neural}
Michael~J. Hartmann and Giuseppe Carleo.
\newblock ``Neural-network approach to dissipative quantum many-body
  dynamics''.
\newblock \href{https://dx.doi.org/10.1103/PhysRevLett.122.250502}{Physical
  Review Letters {\bf 122}, 250502}~(2019).

\bibitem{leclerc2021predicting}
Nima Leclerc.
\newblock ``Predicting dynamics of transmon qubit-cavity systems with recurrent
  neural networks''~(2021).
\newblock  \href{http://arxiv.org/abs/2109.14471}{arXiv:2109.14471}.

\bibitem{flurin2020using}
E.~Flurin, L.~S. Martin, S.~Hacohen-Gourgy, and I.~Siddiqi.
\newblock ``Using a recurrent neural network to reconstruct quantum dynamics of
  a superconducting qubit from physical observations''.
\newblock \href{https://dx.doi.org/10.1103/PhysRevX.10.011006}{Physical Review
  X {\bf 10}, 011006}~(2020).

\bibitem{youssry2020characterization}
Akram Youssry, Gerardo~A. Paz-Silva, and Christopher Ferrie.
\newblock ``Characterization and control of open quantum systems beyond quantum
  noise spectroscopy''.
\newblock \href{https://dx.doi.org/10.1038/s41534-020-00332-8}{npj Quantum
  Information {\bf 6}, 95}~(2020).

\bibitem{lai2021structural}
Zhilu Lai, Charilaos Mylonas, Satish Nagarajaiah, and Eleni Chatzi.
\newblock ``Structural identification with physics-informed neural ordinary
  differential equations''.
\newblock \href{https://dx.doi.org/10.1016/j.jsv.2021.116196}{Journal of Sound
  and Vibration {\bf 508}, 116196}~(2021).

\bibitem{rackauckas2020universal}
Christopher Rackauckas, Yingbo Ma, Julius Martensen, Collin Warner, Kirill
  Zubov, Rohit Supekar, Dominic Skinner, Ali Ramadhan, and Alan Edelman.
\newblock ``Universal differential equations for scientific machine
  learning''~(2020).
\newblock  \href{http://arxiv.org/abs/2001.04385}{arXiv:2001.04385}.

\bibitem{banchi2018modelling}
Leonardo Banchi, Edward Grant, Andrea Rocchetto, and Simone Severini.
\newblock ``Modelling non-{M}arkovian quantum processes with recurrent neural
  networks''.
\newblock \href{https://dx.doi.org/10.1088/1367-2630/aaf749}{New Journal of
  Physics {\bf 20}, 123030}~(2018).

\bibitem{krastanov2020unboxing}
Stefan Krastanov, Kade Head-Marsden, Sisi Zhou, Steven~T. Flammia, Liang Jiang,
  and Prineha Narang.
\newblock ``Unboxing quantum black box models: {L}earning non-{M}arkovian
  dynamics''~(2020).
\newblock  \href{http://arxiv.org/abs/2009.03902}{arXiv:2009.03902}.

\bibitem{heightman2024solving}
Timothy Heightman, Edward Jiang, and Antonio Ac{\'\i}n.
\newblock ``Solving the quantum many-body {H}amiltonian learning problem with
  neural differential equations''~(2024).
\newblock  \href{http://arxiv.org/abs/2408.08639}{arXiv:2408.08639}.

\bibitem{reddy2024data}
Sohail Reddy, Stefanie G{\"u}nther, and Yujin Cho.
\newblock ``{Data-driven characterization of latent dynamics on quantum
  testbeds}''.
\newblock \href{https://dx.doi.org/10.1116/5.0204409}{AVS Quantum Science {\bf
  6}, 033803}~(2024).

\bibitem{cuomo2022scientific}
Salvatore Cuomo, Vincenzo~Schiano Di~Cola, Fabio Giampaolo, Gianluigi Rozza,
  Maziar Raissi, and Francesco Piccialli.
\newblock ``Scientific machine learning through physics--informed neural
  networks: {W}here we are and what’s next''.
\newblock \href{https://dx.doi.org/10.1007/s10915-022-01939-z}{Journal of
  Scientific Computing {\bf 92}, 88}~(2022).

\bibitem{faroughi2022physics}
Salah~A. Faroughi, Nikhil Pawar, Celio Fernandes, Maziar Raissi, Subasish Das,
  Nima~K. Kalantari, and Seyed~Kourosh Mahjour.
\newblock ``Physics-guided, physics-informed, and physics-encoded neural
  networks in scientific computing''~(2022).
\newblock  \href{http://arxiv.org/abs/2211.07377}{arXiv:2211.07377}.

\bibitem{huang2025partial}
Shudong Huang, Wentao Feng, Chenwei Tang, Zhenan He, Caiyang Yu, and Jiancheng
  Lv.
\newblock ``Partial differential equations meet deep neural networks: {A}
  survey''.
\newblock \href{https://dx.doi.org/10.1109/TNNLS.2025.3545967}{IEEE
  Transactions on Neural Networks and Learning Systems {\bf 36},
  13649--13669}~(2025).

\bibitem{kim2021knowledge}
Sung~Wook Kim, Iljeok Kim, Jonghwan Lee, and Seungchul Lee.
\newblock ``Knowledge integration into deep learning in dynamical systems: {A}n
  overview and taxonomy''.
\newblock \href{https://dx.doi.org/10.1007/s12206-021-0342-5}{Journal of
  Mechanical Science and Technology {\bf 35}, 1331--1342}~(2021).

\bibitem{cueto2023thermodynamics}
Elias Cueto and Francisco Chinesta.
\newblock ``Thermodynamics of learning physical phenomena''.
\newblock \href{https://dx.doi.org/10.1007/s11831-023-09954-5}{Archives of
  Computational Methods in Engineering {\bf 30}, 4653--4666}~(2023).

\bibitem{morrison1986paradigm}
Philip~J. Morrison.
\newblock ``A paradigm for joined {H}amiltonian and dissipative systems''.
\newblock \href{https://dx.doi.org/10.1016/0167-2789(86)90209-5}{Physica D:
  Nonlinear Phenomena {\bf 18}, 410--419}~(1986).

\bibitem{grmela1997dynamics}
Miroslav Grmela and Hans~Christian {\"O}ttinger.
\newblock ``Dynamics and thermodynamics of complex fluids. {I}. {D}evelopment
  of a general formalism''.
\newblock \href{https://dx.doi.org/10.1103/PhysRevE.56.6620}{Physical Review E
  {\bf 56}, 6620}~(1997).

\bibitem{ottinger1997dynamics}
Hans~Christian {\"O}ttinger and Miroslav Grmela.
\newblock ``Dynamics and thermodynamics of complex fluids. {II}.
  {I}llustrations of a general formalism''.
\newblock \href{https://dx.doi.org/10.1103/PhysRevE.56.6633}{Physical Review E
  {\bf 56}, 6633}~(1997).

\bibitem{ottinger2005beyond}
Hans~Christian {\"O}ttinger.
\newblock ``Beyond equilibrium thermodynamics''.
\newblock \href{https://dx.doi.org/10.1002/0471727903}{John Wiley \& Sons}.
  ~(2005).

\bibitem{hernandez2021deep}
Quercus Hernandez, Alberto Bad{\'i}as, David Gonz{\'a}lez, Francisco Chinesta,
  and El{\'i}as Cueto.
\newblock ``Deep learning of thermodynamics-aware reduced-order models from
  data''.
\newblock \href{https://dx.doi.org/10.1016/j.cma.2021.113763}{Computer Methods
  in Applied Mechanics and Engineering {\bf 379}, 113763}~(2021).

\bibitem{chinesta2020learning}
Francisco Chinesta, El{\'i}as Cueto, Miroslav Grmela, Beatriz Moya, Michal
  Pavelka, and Martin {\v{S}}{\'\i}pka.
\newblock ``Learning physics from data: {A} thermodynamic interpretation''.
\newblock In Geometric Structures of Statistical Physics, Information Geometry,
  and Learning.
\newblock \href{https://dx.doi.org/10.1007/978-3-030-77957-3_14}{Pages
  276--297}.
\newblock Springer International Publishing~(2021).

\bibitem{hernandez2022thermodynamics}
Quercus Hern{\'a}ndez, Alberto Bad{\'\i}as, Francisco Chinesta, and El{\'\i}as
  Cueto.
\newblock ``Thermodynamics-informed graph neural networks''.
\newblock \href{https://dx.doi.org/10.1109/TAI.2022.3179681}{IEEE Transactions
  on Artificial Intelligence {\bf 5}, 967--976}~(2024).

\bibitem{zhang2022gfinns}
Zhen Zhang, Yeonjong Shin, and George Em~Karniadakis.
\newblock ``G{FINN}s: {GENERIC} formalism informed neural networks for
  deterministic and stochastic dynamical systems''.
\newblock \href{https://dx.doi.org/10.1098/rsta.2021.0207}{Philosophical
  Transactions of the Royal Society A: Mathematical, Physical and Engineering
  Sciences {\bf 380}, 20210207}~(2022).

\bibitem{lee2021machine}
Kookjin Lee, Nathaniel~A. Trask, and Panos Stinis.
\newblock ``Machine learning structure preserving brackets for forecasting
  irreversible processes''.
\newblock Advances in Neural Information Processing Systems {\bf 34},
  5696--5707~(2021).
\newblock  \href{http://arxiv.org/abs/2106.12619}{arXiv:2106.12619}.

\bibitem{gruber2023reversible}
Anthony Gruber, Kookjin Lee, and Nathaniel Trask.
\newblock ``Reversible and irreversible bracket-based dynamics for deep graph
  neural networks''.
\newblock Advances in Neural Information Processing Systems {\bf 36},
  38454--38484~(2023).
\newblock  \href{http://arxiv.org/abs/2305.15616}{arXiv:2305.15616}.

\bibitem{hernandez2023port}
Quercus Hern{\'a}ndez, Alberto Bad{\'\i}as, Francisco Chinesta, and El{\'\i}as
  Cueto.
\newblock ``Port-metriplectic neural networks: thermodynamics-informed machine
  learning of complex physical systems''.
\newblock \href{https://dx.doi.org/10.1007/s00466-023-02296-w}{Computational
  Mechanics {\bf 72}, 553--561}~(2023).

\bibitem{breuer2002theory}
Heinz-Peter Breuer and Francesco Petruccione.
\newblock ``The theory of open quantum systems''.
\newblock
  \href{https://dx.doi.org/10.1093/acprof:oso/9780199213900.001.0001}{Oxford
  University Press}. ~(2007).

\bibitem{lindblad1976generators}
G.~Lindblad.
\newblock ``On the generators of quantum dynamical semigroups''.
\newblock \href{https://dx.doi.org/10.1007/BF01608499}{Communications in
  Mathematical Physics {\bf 48}, 119--130}~(1976).

\bibitem{grabert1982nonlinear}
H.~Grabert.
\newblock ``Nonlinear relaxation and fluctuations of damped quantum systems''.
\newblock \href{https://dx.doi.org/10.1007/BF01314753}{Zeitschrift f{\"u}r
  Physik B Condensed Matter {\bf 49}, 161--172}~(1982).

\bibitem{potts2021thermodynamically}
Patrick~P Potts, Alex Arash~Sand Kalaee, and Andreas Wacker.
\newblock ``A thermodynamically consistent {M}arkovian master equation beyond
  the secular approximation''.
\newblock \href{https://dx.doi.org/10.1088/1367-2630/ac3b2f}{New Journal of
  Physics {\bf 23}, 123013}~(2021).

\bibitem{trushechkin2021unified}
Anton Trushechkin.
\newblock ``Unified {G}orini-{K}ossakowski-{L}indblad-{S}udarshan quantum
  master equation beyond the secular approximation''.
\newblock \href{https://dx.doi.org/10.1103/PhysRevA.103.062226}{Physical Review
  A {\bf 103}, 062226}~(2021).

\bibitem{callen1951irreversibility}
Herbert~B. Callen and Theodore~A. Welton.
\newblock ``Irreversibility and generalized noise''.
\newblock \href{https://dx.doi.org/10.1103/PhysRev.83.34}{Physical Review {\bf
  83}, 34}~(1951).

\bibitem{kubo1966fluctuation}
R~Kubo.
\newblock ``The fluctuation-dissipation theorem''.
\newblock \href{https://dx.doi.org/10.1088/0034-4885/29/1/306}{Reports on
  Progress in Physics {\bf 29}, 255}~(1966).

\bibitem{ottinger2010nonlinear}
Hans~Christian {\"O}ttinger.
\newblock ``Nonlinear thermodynamic quantum master equation: {P}roperties and
  examples''.
\newblock \href{https://dx.doi.org/10.1103/PhysRevA.82.052119}{Physical Review
  A {\bf 82}, 052119}~(2010).

\bibitem{mielke2013dissipative}
Alexander Mielke.
\newblock ``Dissipative quantum mechanics using {GENERIC}''.
\newblock In Recent Trends in Dynamical Systems.
\newblock \href{https://dx.doi.org/10.1007/978-3-0348-0451-6_21}{Pages
  555--585}.
\newblock Springer Basel~(2013).

\bibitem{ottinger2011geometry}
Hans~Christian {\"O}ttinger.
\newblock ``The geometry and thermodynamics of dissipative quantum systems''.
\newblock \href{https://dx.doi.org/10.1209/0295-5075/94/10006}{Europhysics
  Letters {\bf 94}, 10006}~(2011).

\bibitem{mittnenzweig2017entropic}
Markus Mittnenzweig and Alexander Mielke.
\newblock ``An entropic gradient structure for {L}indblad equations and
  couplings of quantum systems to macroscopic models''.
\newblock \href{https://dx.doi.org/10.1007/s10955-017-1756-4}{Journal of
  Statistical Physics {\bf 167}, 205--233}~(2017).

\bibitem{bruning2012parametrizations}
E.~Br{\"u}ning, H.~M{\"a}kel{\"a}, A.~Messina, and F.~Petruccione.
\newblock ``Parametrizations of density matrices''.
\newblock \href{https://dx.doi.org/10.1080/09500340.2011.632097}{Journal of
  Modern Optics {\bf 59}, 1--20}~(2012).

\bibitem{strang2000linear}
Gilbert Strang.
\newblock ``Linear algebra and its applications''.
\newblock Brooks/Cole. ~(2006).
\newblock 4th edition.

\bibitem{paris2004quantum}
Matteo Paris and Jaroslav \v{R}eh{\'a}\v{c}ek, editors.
\newblock ``Quantum state estimation''.
\newblock \href{https://dx.doi.org/10.1007/b98673}{Springer Berlin,
  Heidelberg}. ~(2004).

\bibitem{peng2023qudit}
Zhichao Peng, Daniel Appel{\"o}, N.~Anders Petersson, Mohamad Motamed, Fortino
  Garcia, and Yujin Cho.
\newblock ``Deterministic and {B}ayesian characterization of quantum computing
  devices''~(2023).
\newblock  \href{http://arxiv.org/abs/2306.13747}{arXiv:2306.13747}.

\bibitem{krantz2019quantum}
P.~Krantz, M.~Kjaergaard, F.~Yan, T.~P. Orlando, S.~Gustavsson, and W.~D.
  Oliver.
\newblock ``A quantum engineer's guide to superconducting qubits''.
\newblock \href{https://dx.doi.org/10.1063/1.5089550}{Applied Physics Reviews
  {\bf 6}, 021318}~(2019).

\bibitem{Ramsey1950}
Norman~F. Ramsey.
\newblock ``A molecular beam resonance method with separated oscillating
  fields''.
\newblock \href{https://dx.doi.org/10.1103/PhysRev.78.695}{Physical Review {\bf
  78}, 695--699}~(1950).

\bibitem{kimura2003}
Gen Kimura.
\newblock ``The {B}loch vector for {N}-level systems''.
\newblock \href{https://dx.doi.org/10.1016/S0375-9601(03)00941-1}{Physics
  Letters A {\bf 314}, 339--349}~(2003).

\bibitem{nocedal2006numerical}
Jorge Nocedal and Stephen~J. Wright.
\newblock ``Numerical optimization''.
\newblock \href{https://dx.doi.org/10.1007/978-0-387-40065-5}{Springer, New
  York}. ~(2006).

\bibitem{Place2021}
Alexander P.~M. Place, Lila V.~H. Rodgers, Pranav Mundada, Basil~M. Smitham,
  Mattias Fitzpatrick, Zhaoqi Leng, Anjali Premkumar, Jacob Bryon, Andrei
  Vrajitoarea, et~al.
\newblock ``New material platform for superconducting transmon qubits with
  coherence times exceeding 0.3 milliseconds''.
\newblock \href{https://dx.doi.org/10.1038/s41467-021-22030-5}{Nature
  Communications {\bf 12}, 1779}~(2021).

\end{thebibliography}





\onecolumn
\appendix

\section*{Appendix A: Necessary conditions for $M(\vv)$}

Let $\vv^x$ denote the Bloch representation of the self-adjoint matrix $\rho^x$. From the connection between commutators and the matrix $L(\cdot)$ (see equation~\eqref{eq:L_commutator}), we have
\begin{equation}\label{eq:rho_s_commutator}
    [\rho^s, [\rho^{1-s}, A]]\  \longleftrightarrow\ \widetilde{M}(s)\va,
\end{equation}
where
\begin{equation}\label{eq:M_tilde}
    \widetilde{M}(s) \equiv L^T(\vv^s)L(\vv^{1-s}).
\end{equation}
Since $\rho^s$ and $\rho^{1-s}$ commute, it follows that
\begin{equation}
    [\rho^s, [\rho^{1-s}, A]] = [\rho^{1-s}, [\rho^s, A]],
\end{equation}
from which the symmetry of $\widetilde{M}(s)$ follows immediately.

Note that if $A$ commutes with $\rho$, then $[\rho^s, [\rho^{1-s}, A]]$ for any $s\in (0, 1)$. Thus,
\begin{equation}
    [\rho, A] = 0 \implies \widetilde{M}(s)\va = \mathbf{0}.
\end{equation}
From the relationship between commutators and $L(\cdot)$ in~\eqref{eq:L_commutator}, it follows that
\begin{equation}
    \textrm{null}\left[ L(\vv) \right] \subseteq \textrm{null}\left[\widetilde{M}(s) \right].
\end{equation}
Furthermore, we show in Appendix B that $\widetilde{M}(s)$ is positive semi-definite.

From these properties, it follows that the matrix
\begin{equation}\label{eq:M_v_def}
    M(\vv) \equiv \int_0^1 L^T(\vv^s)L(\vv^{1-s})\dif s,
\end{equation}
must be symmetric positive semi-definite, and also satisfy the null-space condition $\textrm{null}\left[ L(\vv) \right] \subseteq \textrm{null}\left[M(\vv) \right]$. In addition to these two properties, the super-operator $\mathcal{C}_{\rho}'$ takes a particularly simple form when the density matrix corresponds to a pure quantum state, i.e.,
\begin{equation}\label{eq:pure_state_rho}
    \rho = \ket{\psi}\bra{\psi},
\end{equation}
is a rank-one projection onto the span of the quantum state $\ket{\psi}$. In this case, $\rho^x = \rho$ for any positive $x$, so that~\eqref{eq:C_rho_prime} reduces to
\begin{equation}
    \mathcal{C}_\rho'A = -\left[\rho, [\rho, A]\right].
\end{equation}
In terms of the Bloch representation, it follows that when $\rho$ is a pure state, the matrix $M(\vv)$ in~\eqref{eq:M_v_def} takes the form
\begin{equation}
    M(\vv) = L^T(\vv)L(\vv) = -L^2(\vv).
\end{equation}
For $N$-level systems, the set of admissible quantum states is a subset of the hypersphere defined by
\begin{equation}
    r^2 = \|\vv\|^2 \leq \frac{2(N-1)}{N},
\end{equation}
with equality corresponding to pure quantum states.

\section*{Appendix B: Positive semi-definiteness of $\widetilde{M}(s)$}
We prove that the matrix $\widetilde{M}(s)$ defined in~\eqref{eq:M_tilde} is positive semi-definite. Let $A$ and $B$ be two operators with Bloch representations $\va$ and $\vb$ respectively. Assuming at least one of the operators has trace zero, it follows from~\eqref{eq:mat_bloch} and~\eqref{eq:basis_prop} that
\begin{equation}\label{eq:trace}
    \textrm{Tr}(AB) = \frac{1}{2}\va \cdot \vb,
\end{equation}
where $\va \cdot\vb$ is the inner product between $\va$ and $\vb$.

Fix a vector $\vx \in \mathbb{R}^d$, and define the matrix
\begin{equation}
    X \equiv \frac{1}{2}\sum_k x_k\sigma_k,
\end{equation}
which has trace zero. Then from~\eqref{eq:trace} and~\eqref{eq:rho_s_commutator}, we have:
\begin{align}
\begin{split}
    \textrm{Tr}(X[\rho^s, [\rho^{1-s}, X]]) &= \frac{1}{2}\vx^TL^T(s)L(1-s)\vx\\ &= \frac{1}{2}\vx^T\widetilde{M}(s)\vx.
    \end{split}
\end{align}

If we use the eigenvalue decomposition of $\rho$ (with an orthonormal eigenbasis):
\begin{equation}
    \rho = \sum_{j=1}^N r_j\vw_j\vw_j^\dagger,
\end{equation}
then
\begin{equation}
    [\rho^{1-s}, X] = \sum_{j}r_j^{1-s}(\vw_j\vw_j^\dagger X - X\vw_j\vw_j^\dagger),
\end{equation}
and the double commutator satisfies
\begin{align}
    \begin{split}
        [\rho^s, [\rho^{1-s}, X]] &= \sum_{i}r_i(\vw_i\vw_i^\dagger X + X\vw_i\vw_i^\dagger)\\
        - \sum_{ij}&r_i^sr_j^{1-s}((\vw_i^\dagger X\vw_j)\vw_i\vw_j^\dagger + (\vw_j^\dagger X\vw_i)\vw_j\vw_i^\dagger ).
    \end{split}
\end{align}

The trace can be computed using the basis vectors $\vw_k$:
\begin{equation}
   \textrm{Tr}(X[\rho^s, [\rho^{1-s}, X]]) =  \sum_{k}\vw_k^\dagger X[\rho^s, [\rho^{1-s}, X]]\vw_k.
\end{equation}
For fixed $k$ we have:
\begin{align}\label{eq:rho_s_inner}
\begin{split}
    \vw_k^\dagger X[\rho^s, [\rho^{1-s}, X]]\vw_k &= r_k\|X\vw_k\|_2^2 \\
    + \sum_i |\vw_i^\dagger X\vw_k|^2&(r_i - r_i^sr_k^{1-s} - r_i^{1-s}r_k^s).
    \end{split}
\end{align}

For $0\leq s \leq 1$ and any $a,b \geq 0$, we have the standard inequality
\begin{equation}
    a^sb^{1-s} \leq sa + (1-s)b,
\end{equation}
from which it follows that
\begin{equation}\label{eq:r_inequality}
    r_i^sr_k^{1-s} + r_i^{1-s}r_k^s \leq r_i + r_k,
\end{equation}
since $r_i \geq 0$.

Applying~\eqref{eq:r_inequality} to~\eqref{eq:rho_s_inner}, we obtain
\begin{align}
    \begin{split}
        \vw_k^\dagger X[\rho^s, [\rho^{1-s}, X]]\vw_k &\geq r_k\|X\vw_k\|_2^2 - r_k\sum_i |\vw_i^\dagger X\vw_k|^2\\
        &= r_k\left(\|X\vw_k\|_2^2 - \sum_i |\vw_i^\dagger X\vw_k|^2  \right)\\
        &= 0,
    \end{split}
\end{align}
Summing over $k$, we have
\begin{equation}
    \frac{1}{2}\vx^T\widetilde{M}(s)\vx = \textrm{Tr}(X[\rho^s, [\rho^{1-s}, X]]) \geq 0
\end{equation}
showing that $\widetilde{M}(s)$ is positive semi-definite.

\section*{Appendix C: Analytic expression for Nonlinearity $C'_\rho A$} \label{appen:analytical-nonlinearity}

We present here an analytic expression to evaluate the nonlinearity $C'_\rho A$. Consider the spectral decomposition of the density matrix 
\begin{equation}
    \rho = \sum_i^N \lambda_i |\psi_i \rangle \langle \psi_i | = \sum_i^N \lambda_i \Psi_i , 
\end{equation}
%
%
with $\Psi_i = |\psi_i \rangle \langle \psi_i |$ and $\sum_i \lambda_i = 1$.  Then the nonlinear term becomes 
\begin{equation} \label{eq:nonlinear-form1}
\begin{aligned}
    C'_\rho A &= -\int_0^1 \left[ \rho^s, \left[ \rho^{1-s},A \right] \right] ds \\
           &= -\int_0^1 \sum_{i=1}^N \sum_{j=1}^N \lambda^s_i \lambda^{1-s}_j \left[ \Psi_i, \left[ \Psi_j,A \right] \right] ds \\
           &= -\sum_{i=1}^N \sum_{j=1}^N \left[ \Psi_i, \left[ \Psi_j,A \right] \right] \int_0^1  \lambda^s_i \lambda^{1-s}_j ds ,
\end{aligned}
\end{equation}
where the logarithmic mean is defined as 
\begin{equation}
    \Lambda_{ij} = \int_0^1  \lambda^s_i \lambda^{1-s}_j ds = \dfrac{\lambda_i - \lambda_j}{\ln{\lambda_i} - \ln{\lambda_j}},
\end{equation}
%
%
%
with
\begin{equation}
    \lim_{\lambda_j \rightarrow \lambda_i} \dfrac{\lambda_i - \lambda_j}{\ln{\lambda_i} - \ln{\lambda_j}} = \lambda_i.
\end{equation}
Then, letting $\varphi_i$ and $\va$ be the Bloch representations of $\Psi_i$ and $A$, respectively, the nonlinear term reads
\begin{equation}
\begin{aligned}
    C'_\rho A &= -\sum_{i=1}^N \sum_{j=1}^N \Lambda_{ij} L^T(\varphi_i) L(\varphi_j) \va \\
    &= \sum_{i=1}^N \sum_{j=1}^N \Lambda_{ij} L(\varphi_i) L(\varphi_j) \va .
\end{aligned}
\end{equation}
We utilize this expression for generating synthetic data trajectories for the nonlinear thermodynamic master equation in Section \ref{sec:three-level}.

\section*{Appendix D: Experimental Setup and Data Acquisition}
In this work, we used a single-qubit Tantalum-based device mounted at a base temperature of 10 mK. This device is a 2D planar superconducting transmon with Al-AlOx Josephson junction on sapphire substrate \cite{Place2021}. The qubit parameters are estimated with standard characterization protocols, yielding $\omega_{01} = 3.422\,\mathrm{GHz}$.

We used a standard pulse sequence for Rabi measurement to obtain the trajectories used in Section \ref{sec:two-lvl-exp}. We fixed the amplitudes of $p(t)$ and $q(t)$ and varied the pulse lengths to $20\,\mu\mathrm{s}$. The measured data was classified into $|0\rangle$ and $|1\rangle$ states using a Gaussian Mixture Model using \texttt{mixture} in \texttt{sklearn} Python package. The measured populations on $|0\rangle$ and $|1\rangle$ oscillate between 0 and 1 as we apply longer pulses, representing that the Bloch vector rotates around the Bloch sphere vertically. The amplitude ratio between $p(t)$ and $q(t)$ determines the azimuthal angle offset from $x$-axis on the Bloch sphere. Finally, we applied the inverse of a confusion matrix to the measured populations to mitigate state preparation and measurement (SPAM) error. The confusion matrix was given by
\begin{equation}
    \begin{pmatrix}
        0.988875 & 0.010875 & 0.00025 \\
        0.054875 & 0.936375 & 0.00875 \\
        0.016750 & 0.042625 & 0.94063
    \end{pmatrix}.
\end{equation}

To obtain the three Bloch vector components, we applied projection operators before measuring the population. For example, to measure it on Pauli $y$-basis, $\sigma_1$, we applied $\mathrm{RX}(\pi/2)$ gate. A single qubit density matrix can be written as 
\begin{equation}
    \rho=
    \frac{1}{2}\begin{pmatrix}
    1+v_3 & v_1- v_2 i \\
    v_1+ v_2 i & 1-v_3
    \end{pmatrix},
\end{equation}
where $v_j, ~j=1,2,3$ is the Bloch vector component in Pauli bases, $\sigma_j$. When we apply $\mathrm{RX}(\pi/2)$ gate, the density matrix becomes $\rho'= \mathrm{RX}(\pi/2)\rho\,\mathrm{RX}^\dagger (\pi/2)$, resulting in the ground state population $P_y(|0\rangle)=\frac{1}{2}(1+v_2)$. Thus, $v_2 =2P_y(|0\rangle)-1$. Similarly, $v_1 = 2P_x(|0\rangle)-1$ can be obtained by applying $\mathrm{Ry}(\pi/2)$ gate. In superconducting qubits, the default is $z$-basis measurement and $v_3=2P_z(|0\rangle)-1$. 

\section*{Appendix E: Neural Network Training Parameters} 
\captionsetup{type=table}

\captionof{table}{Neural network architecture and training hyperparameters}
\label{tab:nn-training}

\centering
\begin{tabular}{ll}
    \toprule
    \midrule
    Language & Julia \\
    Optimizer, Parameter \\ \quad (used for final training) & \texttt{LBFGS} from \texttt{Optim.jl}, history 10  \\
    Optimizer, Parameter & \texttt{Adam} from \texttt{Optimisers.jl}, learning rate $10^{-1}$ \\
    Continuation step size & $30$ \\
    Maximum optimizer iterations & 2000 for first 50 time points, \\ & \quad 500 for remainder \\
    Hidden MLP layers & $(3,6), (6,9), (9,9)$ \\
    Activation Function & $\tanh$ \\
    Weights initialization & \texttt{setup} from \texttt{Lux.jl} \\
    Initial weight multiplicative scaling & $10^{-6}$ \\
    Tikhonov $\ell^2$ NN weight regularization & $10^{-3}$\\
    Neural ODE integrator & \texttt{Tsit5} from \texttt{OrdinaryDiffEq.jl} \\
    Adjoint method & \texttt{InterpolatingAdjoint(autojacvec = ReverseDiffVJP(true))} \\ & \quad from \texttt{SciMLSensitivity.jl}
    \\ \bottomrule
    
\end{tabular}

\captionof{table}{Physical parameters}
\label{tab:nn-training}

\centering
\begin{tabular}{ll}
    \toprule
    Initialized decay time $T_1$ & 180.0$\; \mu \mathrm{s}$ \\
    Initialized dephase time $T_2$ & 6.0$\;\mu \mathrm{s}$ \\
    Initialized detuning $d$ & 0.0 $\mathrm{rad/\mu s}$\\
    Frequency of rotating frame $\omega_{rot}$ & $(4600\cdot 2\pi -d) \;\mathrm{rad/\mu s}$ \\
    Initialized $\Gamma$ & $\frac{\hbar \omega_{rot}}{2kT}\cdot\mathrm{diag}\left(\frac{1}{T_1},\; \frac{1}{T_1}, \; \frac{2}{T_2}\right)$

\end{tabular}

\end{document}